\documentclass[twocolumn]{aastex63}
\pdfoutput=1 %for arXiv submission
\usepackage{amsmath,amstext,amssymb}
\usepackage[T1]{fontenc}
\usepackage{apjfonts} 
\usepackage{subfigure}
\usepackage{xcolor}
\usepackage{soul}
\usepackage{comment}
\usepackage{graphicx}
\usepackage[normalem]{ulem}
\usepackage{orcidlink}
\usepackage{multirow}

\graphicspath{{figures/}}

\shorttitle{Environment around GW170817 with DESI and standard siren measurement}

\begin{document}

\title{Probing the environment around GW170817 with DESI: insights on galaxy group peculiar velocities for standard siren measurements}

\author[0000-0003-3433-2698]{A.~J.~Amsellem}
\affiliation{Department of Physics, Carnegie Mellon University, 5000 Forbes Avenue, Pittsburgh, PA 15213, USA}

\author{A.~Palmese}
\affiliation{Department of Physics, Carnegie Mellon University, 5000 Forbes Avenue, Pittsburgh, PA 15213, USA}

\author[0000-0002-9540-546X]{K.~Douglass}
\affiliation{Department of Physics \& Astronomy, University of Rochester, 206 Bausch and Lomb Hall, P.O. Box 270171, Rochester, NY 14627-0171, USA}

\author[0000-0002-1081-9410]{C.~Howlett}
\affiliation{School of Mathematics and Physics, University of Queensland, Brisbane, QLD 4072, Australia}

\author[0000-0002-1728-8042]{Juliana~S.~M.~Karp}
\affiliation{SLAC National Accelerator Laboratory, 2575 Sand Hill Road, Menlo Park, CA 94025, USA}

\author{I.~Maga\~{n}a Hernandez}
\affiliation{Department of Physics, Carnegie Mellon University, 5000 Forbes Avenue, Pittsburgh, PA 15213, USA}

\author[0000-0002-2733-4559]{J.~Moustakas}
\affiliation{Department of Physics and Astronomy, Siena University, 515 Loudon Road, Loudonville, NY 12211, USA}

\author[0000-0003-2229-011X]{R.~H.~Wechsler}
\affiliation{Kavli Institute for Particle Astrophysics and Cosmology, Stanford University, Menlo Park, CA 94305, USA}
\affiliation{Physics Department, Stanford University, Stanford, CA 93405, USA}
\affiliation{SLAC National Accelerator Laboratory, 2575 Sand Hill Road, Menlo Park, CA 94025, USA}

\author{J.~Aguilar}
\affiliation{Lawrence Berkeley National Laboratory, 1 Cyclotron Road, Berkeley, CA 94720, USA}

\author[0000-0001-6098-7247]{S.~Ahlen}
\affiliation{Department of Physics, Boston University, 590 Commonwealth Avenue, Boston, MA 02215 USA}

\author[0000-0001-5537-4710]{S.~BenZvi}
\affiliation{Department of Physics \& Astronomy, University of Rochester, 206 Bausch and Lomb Hall, P.O. Box 270171, Rochester, NY 14627-0171, USA}

\author[0000-0001-9712-0006]{D.~Bianchi}
\affiliation{Dipartimento di Fisica ``Aldo Pontremoli'', Universit\`a degli Studi di Milano, Via Celoria 16, I-20133 Milano, Italy}
\affiliation{INAF-Osservatorio Astronomico di Brera, Via Brera 28, 20122 Milano, Italy}

\author{D.~Brooks}
\affiliation{Department of Physics \& Astronomy, University College London, Gower Street, London, WC1E 6BT, UK}

\author[0000-0003-4074-5659]{A.~Carr}
\affiliation{Korea Astronomy and Space Science Institute, 776, Daedeokdae-ro, Yuseong-gu, Daejeon 34055, Republic of Korea}

\author{T.~Claybaugh}
\affiliation{Lawrence Berkeley National Laboratory, 1 Cyclotron Road, Berkeley, CA 94720, USA}

\author[0000-0002-2169-0595]{A.~Cuceu}
\affiliation{Lawrence Berkeley National Laboratory, 1 Cyclotron Road, Berkeley, CA 94720, USA}

\author[0000-0002-4213-8783]{Tamara~M.~Davis}
\affiliation{School of Mathematics and Physics, University of Queensland, Brisbane, QLD 4072, Australia}

\author[0000-0002-1769-1640]{A.~de la Macorra}
\affiliation{Instituto de F\'{\i}sica, Universidad Nacional Aut\'{o}noma de M\'{e}xico,  Circuito de la Investigaci\'{o}n Cient\'{\i}fica, Ciudad Universitaria, Cd. de M\'{e}xico  C.~P.~04510,  M\'{e}xico}

\author[0000-0002-4928-4003]{Arjun~Dey}
\affiliation{NSF NOIRLab, 950 N. Cherry Ave., Tucson, AZ 85719, USA}

\author[0000-0002-5665-7912]{Biprateep~Dey}
\affiliation{Department of Astronomy \& Astrophysics, University of Toronto, Toronto, ON M5S 3H4, Canada}
\affiliation{Department of Physics \& Astronomy and Pittsburgh Particle Physics, Astrophysics, and Cosmology Center (PITT PACC), University of Pittsburgh, 3941 O'Hara Street, Pittsburgh, PA 15260, USA}

\author{P.~Doel}
\affiliation{Department of Physics \& Astronomy, University College London, Gower Street, London, WC1E 6BT, UK}

\author[0000-0002-3033-7312]{A.~Font-Ribera}
\affiliation{Institut de F\'{i}sica d’Altes Energies (IFAE), The Barcelona Institute of Science and Technology, Edifici Cn, Campus UAB, 08193, Bellaterra (Barcelona), Spain}

\author[0000-0002-2890-3725]{J.~E.~Forero-Romero}
\affiliation{Departamento de F\'isica, Universidad de los Andes, Cra. 1 No. 18A-10, Edificio Ip, CP 111711, Bogot\'a, Colombia}
\affiliation{Observatorio Astron\'omico, Universidad de los Andes, Cra. 1 No. 18A-10, Edificio H, CP 111711 Bogot\'a, Colombia}

\author[0000-0001-9632-0815]{E.~Gaztañaga}
\affiliation{Institut d'Estudis Espacials de Catalunya (IEEC), c/ Esteve Terradas 1, Edifici RDIT, Campus PMT-UPC, 08860 Castelldefels, Spain}
\affiliation{Institute of Cosmology and Gravitation, University of Portsmouth, Dennis Sciama Building, Portsmouth, PO1 3FX, UK}
\affiliation{Institute of Space Sciences, ICE-CSIC, Campus UAB, Carrer de Can Magrans s/n, 08913 Bellaterra, Barcelona, Spain}

\author[0000-0003-3142-233X]{S.~Gontcho A Gontcho}
\affiliation{Lawrence Berkeley National Laboratory, 1 Cyclotron Road, Berkeley, CA 94720, USA}
\affiliation{University of Virginia, Department of Astronomy, Charlottesville, VA 22904, USA}

\author{G.~Gutierrez}
\affiliation{Fermi National Accelerator Laboratory, PO Box 500, Batavia, IL 60510, USA}

\author[0000-0002-6550-2023]{K.~Honscheid}
\affiliation{Center for Cosmology and AstroParticle Physics, The Ohio State University, 191 West Woodruff Avenue, Columbus, OH 43210, USA}
\affiliation{Department of Physics, The Ohio State University, 191 West Woodruff Avenue, Columbus, OH 43210, USA}
\affiliation{The Ohio State University, Columbus, 43210 OH, USA}

\author[0000-0002-6024-466X]{M.~Ishak}
\affiliation{Department of Physics, The University of Texas at Dallas, 800 W. Campbell Rd., Richardson, TX 75080, USA}

\author[0000-0003-0201-5241]{R.~Joyce}
\affiliation{NSF NOIRLab, 950 N. Cherry Ave., Tucson, AZ 85719, USA}

\author{R.~Kehoe}
\affiliation{Department of Physics, Southern Methodist University, 3215 Daniel Avenue, Dallas, TX 75275, USA}

\author[0000-0003-3510-7134]{T.~Kisner}
\affiliation{Lawrence Berkeley National Laboratory, 1 Cyclotron Road, Berkeley, CA 94720, USA}

\author[0000-0001-6356-7424]{A.~Kremin}
\affiliation{Lawrence Berkeley National Laboratory, 1 Cyclotron Road, Berkeley, CA 94720, USA}

\author{O.~Lahav}
\affiliation{Department of Physics \& Astronomy, University College London, Gower Street, London, WC1E 6BT, UK}

\author{A.~Lambert}
\affiliation{Lawrence Berkeley National Laboratory, 1 Cyclotron Road, Berkeley, CA 94720, USA}

\author[0000-0003-1838-8528]{M.~Landriau}
\affiliation{Lawrence Berkeley National Laboratory, 1 Cyclotron Road, Berkeley, CA 94720, USA}

\author[0000-0001-7178-8868]{L.~Le~Guillou}
\affiliation{Sorbonne Universit\'{e}, CNRS/IN2P3, Laboratoire de Physique Nucl\'{e}aire et de Hautes Energies (LPNHE), FR-75005 Paris, France}

\author[0000-0003-4962-8934]{M.~Manera}
\affiliation{Departament de F\'{i}sica, Serra H\'{u}nter, Universitat Aut\`{o}noma de Barcelona, 08193 Bellaterra (Barcelona), Spain}
\affiliation{Institut de F\'{i}sica d’Altes Energies (IFAE), The Barcelona Institute of Science and Technology, Edifici Cn, Campus UAB, 08193, Bellaterra (Barcelona), Spain}

\author{V.~Manwadkar}
\affiliation{SLAC National Accelerator Laboratory, 2575 Sand Hill Road, Menlo Park, CA 94025, USA}

\author[0000-0002-1125-7384]{A.~Meisner}
\affiliation{NSF NOIRLab, 950 N. Cherry Ave., Tucson, AZ 85719, USA}

\author{R.~Miquel}
\affiliation{Instituci\'{o} Catalana de Recerca i Estudis Avan\c{c}ats, Passeig de Llu\'{\i}s Companys, 23, 08010 Barcelona, Spain}
\affiliation{Institut de F\'{i}sica d’Altes Energies (IFAE), The Barcelona Institute of Science and Technology, Edifici Cn, Campus UAB, 08193, Bellaterra (Barcelona), Spain}

\author{A.~D.~Myers}
\affiliation{Department of Physics \& Astronomy, University  of Wyoming, 1000 E. University, Dept.~3905, Laramie, WY 82071, USA}

\author[0000-0001-9070-3102]{S.~Nadathur}
\affiliation{Institute of Cosmology and Gravitation, University of Portsmouth, Dennis Sciama Building, Portsmouth, PO1 3FX, UK}

\author[0000-0002-1544-8946]{G.~Niz}
\affiliation{Departamento de F\'{\i}sica, DCI-Campus Le\'{o}n, Universidad de Guanajuato, Loma del Bosque 103, Le\'{o}n, Guanajuato C.~P.~37150, M\'{e}xico}
\affiliation{Instituto Avanzado de Cosmolog\'{\i}a A.~C., San Marcos 11 - Atenas 202. Magdalena Contreras. Ciudad de M\'{e}xico C.~P.~10720, M\'{e}xico}

\author[0000-0003-3188-784X]{N.~Palanque-Delabrouille}
\affiliation{IRFU, CEA, Universit\'{e} Paris-Saclay, F-91191 Gif-sur-Yvette, France}
\affiliation{Lawrence Berkeley National Laboratory, 1 Cyclotron Road, Berkeley, CA 94720, USA}

\author[0000-0002-0644-5727]{W.~J.~Percival}
\affiliation{Department of Physics and Astronomy, University of Waterloo, 200 University Ave W, Waterloo, ON N2L 3G1, Canada}
\affiliation{Perimeter Institute for Theoretical Physics, 31 Caroline St. North, Waterloo, ON N2L 2Y5, Canada}
\affiliation{Waterloo Centre for Astrophysics, University of Waterloo, 200 University Ave W, Waterloo, ON N2L 3G1, Canada}

\author{C.~Poppett}
\affiliation{Lawrence Berkeley National Laboratory, 1 Cyclotron Road, Berkeley, CA 94720, USA}
\affiliation{Space Sciences Laboratory, University of California, Berkeley, 7 Gauss Way, Berkeley, CA  94720, USA}
\affiliation{University of California, Berkeley, 110 Sproul Hall \#5800 Berkeley, CA 94720, USA}

\author[0000-0001-7145-8674]{F.~Prada}
\affiliation{Instituto de Astrof\'{i}sica de Andaluc\'{i}a (CSIC), Glorieta de la Astronom\'{i}a, s/n, E-18008 Granada, Spain}

\author[0000-0001-6979-0125]{I.~P\'erez-R\`afols}
\affiliation{Departament de F\'isica, EEBE, Universitat Polit\`ecnica de Catalunya, c/Eduard Maristany 10, 08930 Barcelona, Spain}

\author[0000-0001-5999-7923]{A.~Raichoor}
\affiliation{Lawrence Berkeley National Laboratory, 1 Cyclotron Road, Berkeley, CA 94720, USA}

\author{G.~Rossi}
\affiliation{Department of Physics and Astronomy, Sejong University, 209 Neungdong-ro, Gwangjin-gu, Seoul 05006, Republic of Korea}

\author[0000-0002-9646-8198]{E.~Sanchez}
\affiliation{CIEMAT, Avenida Complutense 40, E-28040 Madrid, Spain}

\author{D.~Schlegel}
\affiliation{Lawrence Berkeley National Laboratory, 1 Cyclotron Road, Berkeley, CA 94720, USA}

\author{M.~Schubnell}
\affiliation{Department of Physics, University of Michigan, 450 Church Street, Ann Arbor, MI 48109, USA}
\affiliation{University of Michigan, 500 S. State Street, Ann Arbor, MI 48109, USA}

\author[0000-0002-6588-3508]{H.~Seo}
\affiliation{Department of Physics \& Astronomy, Ohio University, 139 University Terrace, Athens, OH 45701, USA}

\author[0000-0002-3461-0320]{J.~Silber}
\affiliation{Lawrence Berkeley National Laboratory, 1 Cyclotron Road, Berkeley, CA 94720, USA}

\author{D.~Sprayberry}
\affiliation{NSF NOIRLab, 950 N. Cherry Ave., Tucson, AZ 85719, USA}

\author[0000-0003-1704-0781]{G.~Tarl\'{e}}
\affiliation{University of Michigan, 500 S. State Street, Ann Arbor, MI 48109, USA}

\author[0000-0001-5381-4372]{R.~Zhou}
\affiliation{Lawrence Berkeley National Laboratory, 1 Cyclotron Road, Berkeley, CA 94720, USA}

\author{The DESI Collaboration}

\begin{abstract}
We present a new measurement of the Hubble constant, $H_0$, following the gravitational wave event GW170817 and Dark Energy Spectroscopic Instrument (DESI) observations. A standard siren measurement with a nearby (luminosity distance $\sim 40 $ Mpc) event such as GW170817 is typically sensitive to the peculiar motion of the host galaxy due to local dynamics. Previous measurements from this event have taken advantage of peculiar velocity measurements of nearby galaxies, including a handful of objects in the galaxy group that the host of the event, NGC 4993, has been associated with. Still, the group's properties and NGC 4993's membership were debated. We present DESI observations of thousands of galaxies in the vicinity of NGC 4993, resulting in 39 group galaxies and a five-fold increase in galaxies compared to previous observations with many of these galaxies contributing to a peculiar velocity measurement. Examining the local dynamics, our observations support the presence of a galaxy group of which NGC 4993 is part with a halo mass of order $\sim$$10^{13}~M_\odot$. Using peculiar velocity measurements from our Fundamental Plane galaxies observations, we find $H_0 =70.9^{+6.4}_{-8.5}$ km s$^{-1}$ Mpc$^{-1}$. In addition, using a peculiar velocity measurement for NGC 4993 from Surface Brightness Fluctuations in Cosmicflows-4 we find $H_0 =73.4^{+3.3}_{-3.9}$ km s$^{-1}$ Mpc$^{-1}$. We study the impact of different galaxy selection criteria on the determination of the peculiar velocity and, in turn, on the $H_0$ measurement.
Our results highlight the importance of multiplexed spectroscopic observations of the environments of gravitational wave events to probe local dynamics, which can ultimately affect standard siren measurements.
\end{abstract}


\section{Introduction}\label{section:intro}
Beyond its most basic significance as a measure of the universe’s ``expansion rate'', the Hubble constant, $H_0$, plays a fundamental role in estimating the age of the universe \citep{freedman_2013}, understanding the velocity components of galaxy motions \citep{t_davis_2014}, and the formulation and verification of dark matter \citep{di_valentino} and dark energy \citep{early_dark_energy} models. Despite its ubiquity in cosmology, there still remains an unresolved discrepancy concerning its value. Currently, two methodologies dominate the landscape of techniques for measuring the Hubble constant: forward modeling of the Cosmic Microwave Background (CMB) and measuring the recessional velocities of standard candles. Using the former method, the Planck Collaboration (\emph{Planck}) have found a value of $H_0 = 67.36 \pm 0.54$ km s$^{-1}$ Mpc$^{-1}$ \citep{planck18}; the latter method, however, has derived a value of $H_0 = 73.04 \pm 1.04$ km s$^{-1}$ Mpc$^{-1}$ \citep{Riess_2022}. Recently, the disagreement between measurements reached a $5\sigma$ level of significance, validating concerns that the tension is trending away from resolution \citep{tension_review, verde, H0_buyers_guide}.

A relatively new, independent method for measuring $H_0$, known as the standard siren method (\citealt{schutz}; see \citealt{2025arXiv250200239P} for a review), could add another competitive measurement to the fray, potentially pointing to a pathway for the tension's resolution. Abstractly, one can ascertain a value for $H_0$ by measuring the luminosity distance, $d_L$, and recessional velocity, $v_{\mathrm{rec}}(z_{\mathrm{rec}})$, from a single object, then solving the approximate Hubble relation, $v_{\mathrm{rec}}(z_{\mathrm{rec}}) \approx H_0 d_L$, for $H_0$. The bright standard siren method applies this reasoning to neutron star mergers with an electromagnetic (or ``bright'') counterpart for which one can measure the luminosity distance via gravitational wave (GW) detection and the recessional redshift via electromagnetic observation. %The standard siren method is therefore conceptually quite simple and does not assume a cosmological history like the CMB-based $H_0$ measurement. 
The standard siren method also harbors an advantage over distance ladder methods: The method relies upon a ``primary'' measure of distance inferred from gravitational waves, obviating the need for calibration of an empirical distance relation as is the case for ``secondary'' distance ladder methods. Thus, concerns regarding distance calibration validity are, for the most part, irrelevant in the case of standard sirens.

The standard siren method can \textit{indirectly} depend upon a distance calibration through the method's connection to peculiar redshift. A galaxy's total redshift, $z_{\mathrm{tot}}$, can be separated into two components, the recessional redshift, $z_{\mathrm{rec}}$, and peculiar redshift, $z_{\mathrm{pec}}$. The peculiar redshift results from the galaxy's local motion rather than the global motion that stems from cosmic expansion. In order to make a standard siren measurement, one must isolate the recessional component of the total redshift by separately measuring the galaxy's peculiar redshift. This peculiar redshift can either be determined via a density-field-based reconstruction of the peculiar redshift  \citep{boruah_2021,mukherjee_2021,blake_turner_vel_recon} or using a distance calibration to directly measure the peculiar redshift. The primary pitfall of the reconstruction method is that it usually relies on a set of assumptions -- e.g. that observed redshifts accurately trace the matter density field \citep{hudson_recon} -- and/or introduces choices within the framework of the $\Lambda$CDM model \citep{graziani}. Conversely, directly measured peculiar redshifts typically involve a distance calibration, which increases uncertainty and potential for systematics biases (see \citealt{standard_siren_speeds} for some examples). In the future the best option for measuring peculiar redshift might be through multimessenger GW sources themselves \citep{Palmese:2020kxn}, but more GW events are needed to perform such measurements. In our standard siren measurement, we choose to measure peculiar velocities directly. We use two types of direct measurement calibrations -- the Fundamental Plane (FP) and Tully-Fisher (TF) relations. Since we implement a direct measurement method, we also probe how different choices in our peculiar velocity analysis systematically affect our $H_0$ result.

GW170817 \citep{gw170817_init}, the only GW event with a confirmed electromagnetic counterpart that has produced a bright standard siren measurement of $H_0$ \citep{gw170817_nature}, was detected in 2017 by the Advanced Laser Interferometer Gravitational-Wave Observatory (LIGO) \citep{2015LIGO} and Advanced Virgo interferometer (Virgo) \citep{Virgo}. This GW observation was quickly followed by a gamma ray burst \citep{gw170817_grb} and lower energy observations of the electromagnetic emission \citep{gw170817_MMA} that led to the identification of NGC 4993 as the galaxy host of GW170817. Further studies indicated that NGC 4993 was potentially a member of a galaxy group \citep[e.g.][]{hjorth_2017,Palmese_2017}. At the time of the GW detection, the peculiar velocity of the NGC 4993's group had already been measured as part of a number of peculiar velocity surveys \citep{crook_ngc4993_pv, crook_ngc4993_pv_erratum, lavaux_ngc4993_pv, makarov_ngc4993_pv, kourkchi_ngc4993_pv}, but subsequent studies further refined these measurements \citep{hjorth_2017, nicolaou, standard_siren_speeds, sbf_ngc4993, cf4}. The NGC 4993 peculiar velocity could be made more precise by combining peculiar velocity measurements from other galaxies in the same group. Yet, all reliable studies only combined peculiar velocity measurements from at most three group galaxies \citep{standard_siren_speeds}. Studies that incorporated more galaxies tended to not thoroughly vet these galaxies' group membership status. It is within this context that we present observations from the Dark Energy Spectroscopic Instrument (DESI; \citealt{Snowmass2013.Levi}) of 14 galaxies near NGC 4993 in order to improve upon previous measurements of the NGC 4993 peculiar velocity and, in turn, the standard siren measurement of $H_0$.

The outline of the paper is as follows. In Section \ref{section:data}, we describe the target selection process, our peculiar velocity galaxy samples, and the gravitational wave data. In Section \ref{section:methods}, we parameterize the posterior formulation for a bright standard siren $H_0$ measurement, parameterize our implementation of the Fundamental Plane and Tully-Fisher relations, and describe the zero-point calibration procedure for these relations. In Section \ref{section:results}, we present our measured values of the group peculiar velocity and the Hubble constant while performing a number of measurements relating to the group's virialization and relaxation. In Sections \ref{section:Discussion} and \ref{section:Conclusion}, we provide a short discussion and summary of our results. All uncertainties reported are $1\sigma$, and we report the median along with the 68\% highest density interval when quoting results from our posteriors.

\section{Data}\label{section:data}

\subsection{DESI observations}\label{subsection:desi_obs}
To measure the peculiar velocity of NGC 4993, we use spectroscopy from the Dark Energy Spectroscopic Instrument (DESI) and photometry from the Dark Energy Camera Legacy Survey (DECaLS) in the 10th data release (DR10) of the DESI Legacy Survey \citep{legacy_survey}. DESI utilizes the Mayall four-meter telescope located at Kitt Peak National Observatory \citep{DESI2022.KP1.Instr} to take spectroscopic observations primarily in the northern portion of the sky. DESI consists of 10 spectrographs each connected to 500 robotically operated optical fibers, allowing for simultaneous observation of 5,000 unique sky locations in a region with a $\sim$$1.63^{\circ}$ radius \citep{DESI2016b.Instr, FocalPlane.Silber.2023, Corrector.Miller.2023, FiberSystem.Poppett.2024}. The primary role of DESI is to probe the universe's expansion history in order to better characterize the nature of dark energy \citep{Snowmass2013.Levi} with recent results already yielding exciting constraints on the dark energy equation of state \citep{DESI2024.VII.KP7B, DESI.DR2.BAO.cosmo}. To this end, DESI has measured over $50$ million redshifts -- surpassing its five-year goal of $40$ million redshifts \citep{DESI2016a.Science}. In this work, redshifts are measured via the standard spectroscopic reduction pipeline (desispec/0.59.0) that relies upon matching spectra to templates \citep{Spectro.Pipeline.Guy.2023}. Since NGC 4993 lies outside the DESI footprint, our observations signify a departure from the typical survey operations schedule \citep{Schlafly_survey_ops} as part of a tertiary target program \citep{myers_target_selection} to specifically characterize the peculiar velocity and environment of NGC 4993. Our peculiar velocity measurement is rooted in FP (Section \ref{subsection:fp_method}) and TF (Section \ref{subsection:tf_method}) peculiar velocity calibrations, which were constructed using data from the DESI Survey Validation observations \citep{DESI2023a.KP1.SV} as found in the DESI Early Data Release \citep{DESI2023b.KP1.EDR} as well as observations from the DESI Year 1 (Y1) observing run \citep{DESI2025.KP2.DR1}, which corresponds to Data Releas 1 (DR1). This work marks one of the first instances in which these peculiar velocity calibrations were used to measure a galaxy's peculiar velocity outside of the fiducial DESI sample.

As mentioned above, the measurement of the NGC 4993 peculiar velocity requires measurements from other galaxies in the NGC 4993 group. However, the task of observing galaxies in the NGC 4993 group presents an issue: We seek to observe TF and FP galaxies within the group, but until they are observed we cannot determine a) their membership in the group \emph{a priori} or b) whether or not their peculiar velocities can be measured. The simplest and most comprehensive resolution to this issue is to observe as many galaxies that could potentially be used in our group peculiar velocity measurement. Without a multi-object spectrograph with thousands of fibers like DESI, completing this task in a reasonable amount of time would be unfeasible. With this task in mind, we culled a subset of 923 objects from the targets selected for the DESI Peculiar Velocity Survey and observed them in June of 2023. In anticipation of a second set of observations in June of 2024, we added 52 galaxies to our original set of targets based upon visual inspection of our photometric observations. When considering our full list of targets, 828 targets were classified as FP targets while 147 were classified as TF targets based upon the photometric selection criteria outlined in \citep{target_selection}. The 52 galaxies added for the second observing run were categorized as TF or FP by visual inspection and color characterization -- bluer indicating a TF galaxy and redder indicating an FP galaxy -- with the more quantitative selection criteria described below being invoked after observation. In addition to our PV targets, the DESI LOW-Z Survey \citep{lowz} simultaneously submitted 47,071 galaxy targets near NGC 4993 for observation as part of an unrelated scientific analysis; we have included these galaxies as part of our group membership determination in Section \ref{subsection:cluster_mem}. After acquiring DESI spectroscopic redshifts for both the LOW-Z and our PV targets, we focus on galaxies with redshifts in the regime $0.008<z_{\mathrm{tot}}<0.012$. This regime is close to the redshift of NGC 4993, but we justify this choice further in \ref{subsection:cluster_mem}. Invoking this redshift cut, we find 17 potential TF galaxies and nine potential FP galaxies to be used in our PV analysis. All targets that were observed by DESI are shown in Figure~\ref{fig:{footprint}}. In this figure, we specify general targets (blue), TF group targets (green), FP group targets (orange), and non-PV group targets (red).

In order to measure its peculiar velocity, each TF galaxy was observed at its center and at a distance from the center along its semi-major axis of 40\% of $R_{26}$, where $R_{26}$ is the radius within which the magnitude density is 26 mag arcsec$^{-2}$ in the $r$-band. This very specific off-center requirement is chosen to match the DESI DR1 TF Peculiar Velocity Survey calibration radius \citep{douglass_tfr_Y1}. We determine $R_{26}$ for each galaxy using the \texttt{legacyhalos} software,\footnote{\url{https://github.com/moustakas/legacyhalos}} which determines a galaxy's isophotes and position angle via an ellipse-fitting procedure. Using these fitted parameters, we determine two off-center target observations for each TF galaxy along its major axis.

We also use \texttt{legacyhalos} more generally for measuring and deriving uncertainties on all photometric quantities in this study. We use the output from the \texttt{tractor} software\footnote{\url{https://github.com/dstndstn/tractor}} (embedded in the \texttt{legacyhalos} package) for FP-related photometry \citep{lang_tractor} and \texttt{legacyhalos}' ellipse-fitting techniques for TF-related photometry. The difference in source of photometry is reflective of the DESI FP and TF calibrations, which use \texttt{tractor} and \texttt{legacyhalos} ellipse-fitting, respectively.

\subsection{GW and Afterglow data}\label{subsection:gw_obs}
As part of our standard siren $H_0$ measurement, we require the posterior of the GW170817 luminosity distance and inclination angle. Although the strain data for GW170817 was made public as a part of GWTC-1 \citep{gwtc1}, we found that estimating GW parameter posteriors directly from the strain data to be computationally prohibitive and excessively time-consuming. Instead, we glean information regarding the merger's luminosity distance and inclination angle from the public high-spin GW170817 posterior samples\footnote{\url{https://dcc.ligo.org/LIGO-P1800061/public}} \citep{gw170817_posterior}. Using a twelve-component gaussian mixture model from \texttt{scikit-learn}, we reconstruct a 2D posterior surface from the luminosity distance and inclination angle samples.

We also consider inclination angle information derived from the merger's afterglow as part of our measurement. Following \citet{palmese_ag}, we fit a twelve-component gaussian mixture model from \texttt{scikit-learn} to reconstruct a 1D posterior surface of the inclination angle based upon the posterior samples derived using the \texttt{JetFit} package \citep{wu_2018}. The afterglow data is compiled in \citet{Hajela_2022} with data observed by Chandra x-ray Observatory, Karl G. Jansky Very Large Array, Hubble Space Telescope, and MeerKAT \citep{hallinan_2017,alexander_2017,mooley_2018a,margutti_2017,margutti_2018,dobie_2018,alexander_2018,mooley_2018b,hajela_2019,fong_2019}. Both the GW and afterglow posteriors are used as a part of our posterior determination of $H_0$ (see Section \ref{subsection:posterior}).

\begin{figure*}[htbp]
  \centering
    \includegraphics[width=\linewidth]{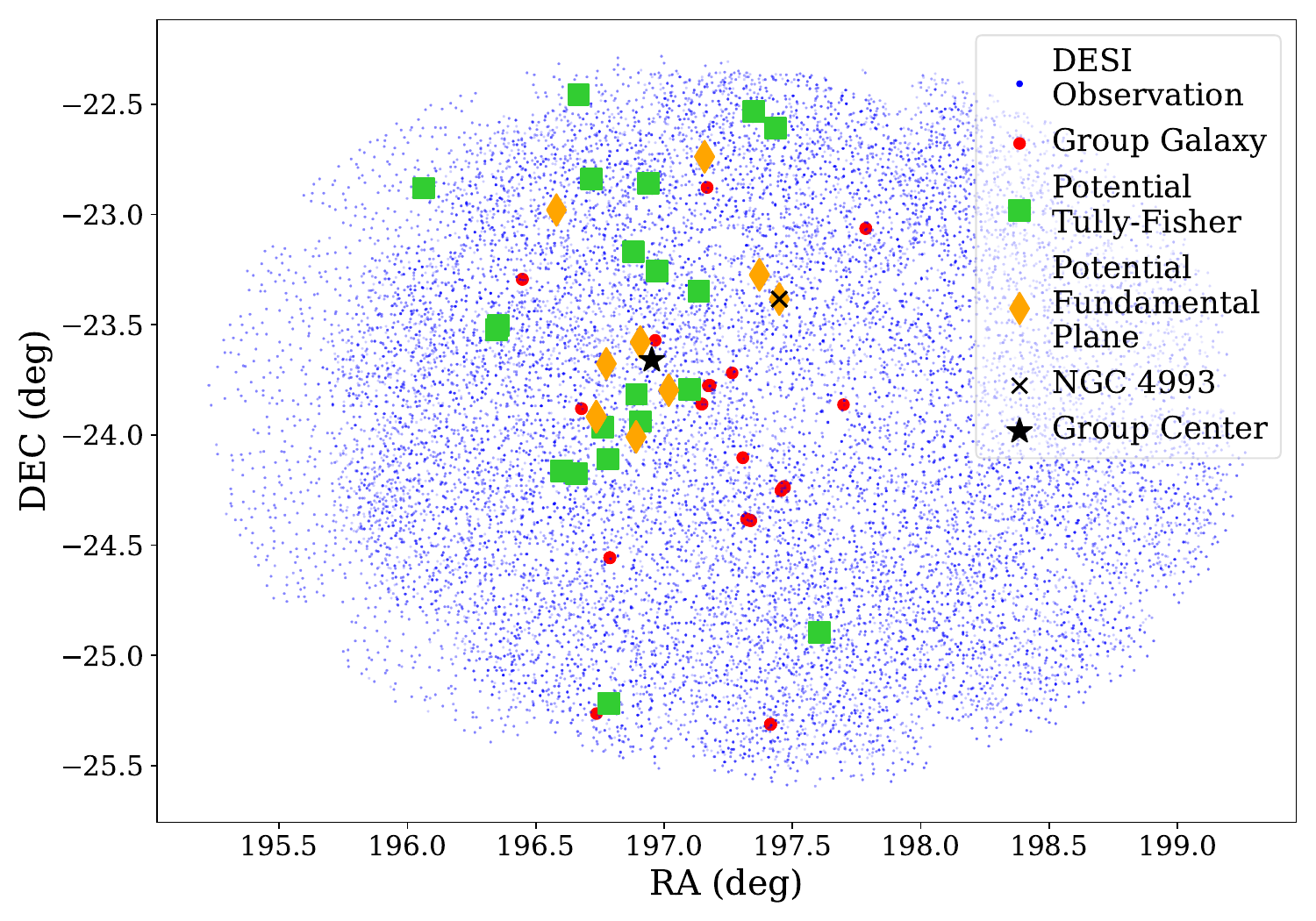}
    \caption{\textbf{On-sky depiction of DESI observations.} Each blue circle indicates one of the 47,071 observations -- as targeted by our PV Survey and the DESI LOW-Z Survey -- in the on-sky vicinity of NGC 4993. Each red circle indicates a galaxy in the same galaxy group as NGC 4993 that is neither an FP nor a TF galaxy. Our TF group targets are demarcated by green squares, and the FP group targets are demarcated by yellow diamonds. One of the FP galaxies is NGC 4993 itself, indicated by a black cross. The approximate center of the galaxy group is designated by a black star.}
    \label{fig:{footprint}}
\end{figure*}

\section{Methods}\label{section:methods}
\subsection{Standard siren posterior}\label{subsection:posterior}
We now turn to the construction of a Bayesian probability model for the posterior distribution of $H_0$ as has been similarly derived in previous work \citep{gw170817_nature}. In a departure from the more common posterior definition in terms of peculiar velocity, we choose to work in terms of log-distance ratio, $\eta$, defined as 
\begin{equation}\label{eq:log_distance}
\begin{aligned}
\eta(z_{\mathrm{tot}}, z_{\mathrm{cmb}}, d_L, H_0) &= \mathrm{log}_{10} \left[\frac{d_c(z_{\mathrm{cmb}}, H_0)}{d_c({z_{\mathrm{rec}}, H_0})}\right]\\
                        &= \mathrm{log}_{10} \left[(1 + z_{\mathrm{rec}}) \cdot \frac{d_c(z_{\mathrm{cmb}}, H_0)}{d_L(z_{\mathrm{cmb}}, H_0)} \right]\\
                        &\approx \mathrm{log}_{10} \left[(1 + z_{\mathrm{tot}}) \cdot \frac{d_c(z_{\mathrm{cmb}}, H_0)}{d_L(z_{\mathrm{cmb}}, H_0)} \right]\\
\end{aligned}
\end{equation}
where $z_{\mathrm{tot}}$ is the total redshift in the heliocentric frame, $z_{\mathrm{cmb}}$ is the total redshift in the CMB frame, $d_c(z, H_0)$ is the comoving distance as a function of a redshift and a Hubble constant value, and $d_L$ is the luminosity distance derived from the GW event waveform. %Since the peculiar redshift, $z_{\mathrm{pec}}$, satisfies
%\begin{equation}
%(1 + z_{\mathrm{tot}}) = (1+z_{\mathrm{rec}}) \times (1+z_{\mathrm{pec}}),
%\end{equation}
%the final line of Equation \ref{eq:log_distance} amounts to the assumption that $\mathrm{log}_{10}(1+z_{\mathrm{pec}})$ is negligible, allowing for the replacement of $z_{\mathrm{rec}}$ with $z_{\mathrm{tot}}$.
Log-distance ratios can be considered preferable to peculiar velocities since the former are directly derived from observables, require few simplifying approximations, and possess statistical uncertainties that are reasonably approximated by a Gaussian distribution \citep{standard_siren_speeds}. Moreover, in the redshift regime of NGC 4993, log-distance ratios can be converted to peculiar velocities via the approximation derived in \citep{watkins_feldman},
\begin{equation}\label{eta_to_pv}
\begin{aligned}
v_{\mathrm{pec}} \approx \frac{c z_{\mathrm{mod}}}{(1 + z_{\mathrm{mod}})} \mathrm{ln}(10) \cdot \eta
\end{aligned}
\end{equation}
where $z_{\mathrm{mod}}$ is an expansion of $z_{\mathrm{cmb}}$ in terms of deceleration parameter, $q_0$, and jerk, $j_0$, such that 
\begin{equation}\label{eq:zmod}
\begin{aligned}
z_{\mathrm{mod}} = z_{\mathrm{cmb}} \left[1 + \frac{1}{2}(1-q_0) z_{\mathrm{cmb}} - \frac{1}{6}(1 - q_0 - 3q_0^2 + j_0) z_{\mathrm{cmb}}^2\right].
\end{aligned}
\end{equation}
This redshift modification has been added in order to account for cosmic acceleration \citep{t_davis_2014}. We assume $q_0 \approx 1 - \frac{3}{2} \Omega_m = -0.53551$, using \emph{Planck}'s measured matter density, $\Omega_m$, \citep{planck18} and neglecting the small density contributions from curvature and radiation. We assume $j_0=1$, which is the value it takes in Flat-LCDM cosmology. We note that our peculiar velocity results are robust to alternative values of $q_0$ and $j_0$.

With this understanding of log-distance ratio in hand, we can now formulate a posterior for $H_0$. We begin by writing down the conditional probability of observing, $X$, the data associated with the merger, given $\theta$, the set of true parameters that describe the merger and the universe. We define $X = \{X_{\mathrm{GW}}, X_{\mathrm{EM}}, \eta_{\mathrm{obs}}\}$ where $X_{\mathrm{GW}}$ represents the GW data, $X_{\mathrm{EM}}$ represents the EM data, and $\eta_{\mathrm{obs}}$ is the observed log-distance ratio of the host group. We also define $\theta = \{d_L, z_{\mathrm{tot}}, \mathrm{cos}(\iota), H_0\}$ where $d_L$ is the group luminosity distance, $z_{\mathrm{tot}}$ is the total group redshift in the heliocentric frame (taken to be the redshift of the brightest galaxy in the group), and $\mathrm{cos}(\iota)$ is the cosine of the angle between our line of sight to the merger and the axis of rotation of the binary neutron star system (i.e. the inclination angle). We then separate $p(X|\theta)$ into three independent probabilities 
\begin{equation}\label{eq:p(x|theta)}
\begin{aligned}
    p(X|\theta) = p(X_{\mathrm{GW}}|d_L, \mathrm{cos}(\iota)) p(X_{\mathrm{EM}}|z_{\mathrm{tot}})  p(\eta_{\mathrm{obs}}|z_{\mathrm{tot}}, d_L, H_0).
\end{aligned}
\end{equation}
Invoking Bayes' theorem and marginalizing over all parameters in $\theta$ besides $H_0$, we find
\begin{equation}\label{eq:H0_post}
\begin{aligned}
    p(H_0|X) \propto p(H_0) \int &\mathrm{d}d_L \: \mathrm{d}z_{\mathrm{tot}} \: d\mathrm{cos}(\iota) \: p(X_{\mathrm{GW}}|d_L, \mathrm{cos}(\iota))\\
    &\times p(X_{\mathrm{EM}}|z_{\mathrm{tot}}) p(\eta_{\mathrm{obs}}|z_{\mathrm{tot}}, d_L, H_0)\\ &\times p(d_L) p(z_{\mathrm{tot}}) p(\mathrm{cos}(\iota))
\end{aligned}
\end{equation}
where $p(H_0)$ is a uniform $H_0$ prior between $U [20,120] ~\mathrm{km~ s^{-1}~Mpc^{-1}}$, $p(z_{\mathrm{tot}})$ is $U[0.0077, 0.0125]$, and $p(\mathrm{cos}(\iota))$ is $U[0.3, 0.6]$. We have chosen to ignore both the evidence from Bayes' theorem and a term to account for selection effects since both would only amount to a constant factor in the $H_0$ posterior in the case of a measurement from only GW170817 \citep{gw170817_nature}. 
The distributions $p(X_{\mathrm{GW}}|d_L, \mathrm{cos}(\iota)) p(d_L)$ and $p(X_{\mathrm{EM}}|z_{\mathrm{tot}})$ are derived from posterior samples as described in Section \ref{subsection:gw_obs}.  Lastly, we take $p(\eta_{\mathrm{obs}}|z_{\mathrm{tot}}, d_L, H_0)$ to be a Gaussian with a mean of $\eta_{\mathrm{obs}}$ and standard deviation $\sigma_{\eta_{\mathrm{obs}}}$ with these values being measured from our DESI galaxy observations. The log-distance ratio domain of this Gaussian is constructed from parameters $z_{\mathrm{tot}}$, $d_L$, $H_0$ using Equation \ref{eq:log_distance}.

We use Equation \ref{eq:H0_post} for our fiducial measurement of the $H_0$ posterior, but we also modify Equation \ref{eq:H0_post} to examine alternative characterizations of the standard siren posterior. One alteration changes $p(\eta_{\mathrm{obs}}|z_{\mathrm{tot}}, d_L, H_0)$ to $p(\eta_{\mathrm{obs}}|z_{\mathrm{tot}}, d_L, H_0, s)$ where $s$ is a ``smoothing kernel'' used to assign a weight to each galaxy's log-distance ratio considered for the PV measurement. Incorporating $s$ allows for a down-weighting of galaxies farther from the kernel center that are less likely to be group members. Since we are careful in our treatment of group membership, smoothing kernel considerations are unnecessary for our fiducial posterior analysis, but we include it in as an ancillary analysis to see if it can significantly alter our results. Under the smoothing kernel conception of the $H_0$ posterior, the weight for the $i$th galaxy is parameterized as 
\begin{equation}
    w_i = e^{\frac{2 s^2}{\Delta v_i}}
\end{equation}
where $\Delta v_i$ is the difference in CMB-frame velocity of the $i$th galaxy and kernel center velocity in the supergalactic Cartesian frame. With this parametrization of $p(H_0|X)$, we marginalize over $s$ as was first done in \citet{nicolaou}.

A second alteration to our parametrization of the $H_0$ posterior accounts for differences between group and NGC-4993-specific quantities. When determining $\eta$ via Equation \ref{eq:log_distance}, one ought to use the redshift of the merger. Since we cannot measure that redshift, we can use the redshift of the merger's galaxy host as an approximation of the actual merger redshift. In our analysis, we take this approximation one step further: we approximate the redshift of the host galaxy to be the same as the redshift of the group as a whole. It should be noted that the usage of the group redshift rather than the galaxy host redshift denotes a conceptual consistency with the usage of the group PV rather than the galaxy host PV. This extra approximation, however, comes at a systematic cost: There is a fixed -- albeit minute -- difference between the group center and galaxy host redshifts. Thus, our analysis assumes a merger redshift that is known to be systematically off from its true value. Since we surmise that the merger occurs within the group, this systematic offset ought to be so minute that it will not affect our $H_0$ posterior. To examine this hypothesis, we model this offset as a free perturbative term to $d_L$ in Equation \ref{eq:log_distance} such that the host is allowed to be at a different distance than the group. Heuristically, we set a uniform prior on this parameter, allowing it to vary within plus or minus one virial radius of the group.
% We parameterize the term as the difference between the comoving distance of the group center and that of the galaxy host. We note that this perturbation is calculated based on the \textit{total} group and host redshifts, which amounts to the assumption that the measured comoving distance between the group center and the host is not sullied by a significant difference between the group and host peculiar velocities.
% \begin{equation}\label{eq:dL_pert}
% \begin{aligned}
% \eta(z_{\mathrm{tot}}, d_L, H_0) \approx \mathrm{log}_{10} \left[\frac{d_c(z^{\mathrm{gr}}_{\mathrm{cmb}}, H_0)}{\frac{d_L}{(1 + z^{\mathrm{gr}}_{\mathrm{tot}})} + d_c(z^{\mathrm{gr}}_{\mathrm{cmb}}, H_0) - d_c(z^h_{\mathrm{cmb}}, H_0)}  \right]\\
% \end{aligned}
% \end{equation}
% where $z^{\mathrm{gr}}_{\mathrm{cmb}}$ and $z^{\mathrm{gr}}_{\mathrm{tot}}$ represent the CMB-frame and heliocentric-frame total group redshifts, $z^h_{\mathrm{cmb}}$ represents the CMB-frame total galaxy host redshift, and $d_c(z^{\mathrm{gr}}_{\mathrm{cmb}}, H_0) - d_c(z^h_{\mathrm{cmb}}, H_0)$ represents the systematic offset between the comoving distance of the group center and the galaxy host. We note that this perturbation is calculated based on the \textit{total} group and host redshifts, which amounts to the assumption that the measured comoving distance between the group center and the host is not sullied by a significant difference between the group and host peculiar velocities.
To recapitulate, the smoothing kernel and $d_L$ perturbation forms of the $H_0$ posterior will be examined, but our fiducial $H_0$ measurement makes use of the posterior outlined in Equation \ref{eq:H0_post}.

We approximate the integral in Equation \ref{eq:H0_post} by taking many prior-constrained samples of the parameters in $\theta$, evaluating the likelihood for each set of samples, and constructing the posterior from these likelihood evaluations. This procedure is implemented using the static nested sampling code found in \texttt{dynesty} \citep{dynesty_speagle, dynesty_koposov, dynesty_skilling_1, dynesty_skilling_2} via the Bayesian inference Python package \texttt{bilby} \citep{bilby_paper}.

\subsection{Fundamental Plane}\label{subsection:fp_method}
The Fundamental Plane is an empirical relationship for elliptical galaxies between three galaxy observables -- the galaxy's effective radius, $R_e$; the mean surface brightness, $I_e$; and central velocity dispersion, $\sigma_0$. As mentioned in Section \ref{section:intro}, the measurement of $\eta_{\mathrm{obs}}$, which appears in Equation \ref{eq:H0_post}, necessitates the measurement of log-distance ratios from NGC 4993 group galaxies. Following \citep{fp_sdss} and \citep{ross_fp_Y1}, we outline how the three observables quantities are themselves calculated from more fundamental photometric and spectroscopic observables.

The effective radius can be calculated via the following relation
\begin{equation}\label{eq:eff_rad}
R_e = r \sqrt{\frac{b}{a}} \cdot \frac{d_c(z_{\mathrm{cmb}})}{1+z_{\mathrm{helio}}}
\end{equation}
where $r$ is the half-light radius in arcseconds, $\frac{b}{a}$ is the axis ratio that characterizes a fitted ellipse of the galaxy, $d_c(z_{\mathrm{cmb}})$ is the comoving distance as calculated in from the galaxy's total CMB-frame redshift, and $z_{\mathrm{helio}}$ is the total heliocentric-frame redshift. $r$ and $\frac{b}{a}$ are determined by performing ellipse-fitting photometry in a similar fashion to that described in \citet{sga}. It is important to note that using Equation \ref{eq:eff_rad}, one can calculate $R_e$ in units of Mpc $h^{-1}$ by calculating $d_c(z_{\mathrm{cmb}})$ in those units; thus, $R_e$ is a quantity that is $H_0$-independent.

Moving to the next axis of the FP, we define the mean surface brightness as the fraction of observed brightness contained within the area of an ellipse parameterized by $r$ and $\frac{b}{a}$,
\begin{equation}\label{eq:surf_bright}
I_e = \frac{10^{0.4[M^r_{\odot} - m_r - 0.85 z_{\mathrm{cmb}} + k_r]}}{2 \pi r^2 \frac{b}{a}} (1 + z_{\mathrm{helio}})^4,
\end{equation}
where $M^r_{\odot}=4.65$ sets the conventional brightness scale to the Sloan Digital Sky Survey $r$-band absolute magnitude of the Sun \citep{willmer}, $m_r$ is the galaxy's $r$-band apparent magnitude (corrected for Milky Way extinction), $r$ is the half-light radius computed in radians, and $k_r$ is a k-correction factor to the apparent magnitude as described in \citet{kcorr}. Equation \ref{eq:surf_bright} also contains two other necessary magnitude corrections: the $(1+z_{\mathrm{helio}})^4$ factor corrects for cosmological surface brightness dimming \citep{tolman} while the $0.85 z_{\mathrm{cmb}}$ factor corrects for a surface brightness bias at higher redshifts \citep{Bernardi}. We tested an alternative $1.1 z_{\mathrm{cmb}}$ correction factor in concordance with \citet{ross_fp_Y1}, but found the effects at the low redshift of NGC 4993 to be negligible.

The final FP observable, the central velocity dispersion $\sigma_0$ is defined as the standard deviation of the velocities of stars contained within a galaxy. The velocity dispersion is different from the other FP observables in that it is directly measured from a galaxy spectrum using the Penalized PiXel-Fitting (pPXF) python package. To measure velocity dispersion, pPXF fits an observed galaxy spectrum to a convolution of stellar templates and a Line-of-Sight Velocity Distribution where the velocity dispersion characterizes the broadening of the stellar emission and absorption peaks \citep{ppxf}. For the fitting, we use stellar templates from the Indo-U.S. Coud{\'e} Feed Spectral Library \citep{indous_templates}. We then convert this fitted velocity dispersion of the entire galaxy, $\sigma$, to a velocity dispersion of the galaxy's central region, $\sigma_0$, via the equation \citep[similar to the one derived in][]{jorgensen}
\begin{equation}
    \sigma_0 = \sigma \left(\frac{r \sqrt{b/a}}{8 \theta_{\mathrm{ap}}}\right)^{-0.04}
\end{equation}
where $\theta_{\mathrm{ap}}=0.75$ arcseconds is the radius of a DESI fiber.

We have now demonstrated how the three FP parameters can be derived from basic photometric and spectroscopic observables. These three observables are also related to one another via the FP relation. Using this relationship, one can arrive at a calibrated value for the effective radius $R^{\mathrm{cal}}_e$:
\begin{equation}\label{eq:fund_plane}
\mathrm{log}_{10}(R^{\mathrm{cal}}_e) = a \mathrm{log}_{10}(\sigma_0) + b \mathrm{log}_{10}(I_e) + c
\end{equation}
where $a$ and $b$ are the relation's calibrated slope parameters while $c$ is the relation's calibrated zero point. DESI has already produced measurements of these calibration parameters using data from the survey validation phase of operations \citep{said_fpr_edr} and the DR1 data release \citep{ross_fp_Y1}, which we use to calculate the effective radius. Equation \ref{eq:fund_plane} represents a second method of calculating effective radius in addition to the direct photometric method shown in Equation \ref{eq:eff_rad}. $R_e$ in Equation \ref{eq:eff_rad} reflects an effective radius that is either over- or under-estimated due to the presence of peculiar velocities. On the other hand, $R^{\mathrm{cal}}_e$ in Equation \ref{eq:fund_plane} reflects the true effective radius since the relationship filters out the random spread of peculiar velocities by calibrating with a large sample of galaxies. The ratio of these two radii therefore contains information about the galaxy's peculiar velocity. In fact, by converting these radii into angular diameter distances, one can show that
\begin{equation}
    \eta_{\mathrm{gal}, \mathrm{FP}} = \mathrm{log}_{10}\left(\frac{R_e}{R^{\mathrm{cal}}_e}\right)
\end{equation}
where $\eta_{\mathrm{gal}, \mathrm{FP}}$ is the log-distance ratio of a single galaxy (as opposed to $\eta_{\mathrm{obs}}$ above which is the log-distance ratio of NGC 4993, derived by combining the log-distance ratios of a number of group galaxies). Thus, from a combination of photometry, spectroscopy, and the Fundamental Plane calibration, we can derive the log-distance ratio for the observed elliptical galaxies in the NGC 4993 group.

To derive an uncertainty on $\eta_{\mathrm{gal}, \mathrm{FP}}$, we use a log-likelihood minimization technique. We closely follow the minimization procedure outlined in \citet{fp_sdss}; we point the reader to that work for a more thorough description of this procedure. The uncertainty on $\eta_{\mathrm{gal}, \mathrm{FP}}$ derives from two sources: intrinsic scatter on the FP relation and photometric measurement error. The intrinsic scatter is measured on each the three FP axes as part of the FP calibration procedure \citep{ross_fp_Y1} while the photometric errors are derived using \texttt{tractor}. From these uncertainties, we construct a covariance matrix for each galaxy. We then perform a log-likelihood minimization over a range of log-distance ratios using our measured FP observables, FP calibration parameters, and the covariance matrix. The minimization provides a posterior distribution for $\eta_{\mathrm{gal}, \mathrm{FP}}$ of each galaxy. We then fit the uncertainty on each $\eta_{\mathrm{gal}, \mathrm{FP}}$ posterior with a skew-normal distribution. Lastly, we note that based on Figure 9 in \citep{fp_sdss}, we determine that the selection biases for galaxies at or around the distance of NGC 4993 (i.e. $\sim40$ Mpc) should be negligible and are therefore ignored.

\subsection{Tully-Fisher}\label{subsection:tf_method}
For spiral galaxies in the NGC 4993 group, one can also derive log-distance ratios using a TF calibration. Unlike the FP relation, the TF relation consists of only two axes -- one from the galaxy's rotational velocity and the other from its absolute magnitude. We will outline how these quantities are derived from more fundamental observables and then explain how to derive log-distance ratios using the TF relation. The derivation of the TF observables closely follows a similar procedure described in \citet{douglass_tfr_Y1} since our measured TF observables must conform to those used in the TF calibration sample.

One can determine the rotational velocity of a galaxy by observing the difference in redshift at the center and edge of a galaxy. As in the DESI DR1 TF calibration, we define the ``galaxy edge'' to be the point(s) along the semi-major axis that lie at a distance of $0.4R_{26}$ from the center (with this quantity already being defined in Section \ref{subsection:desi_obs}). In instances when we have multiple off-center redshift measurements, we determine the rotational velocity using a weighted average of the differences between the off-center redshifts and center redshift. For galaxies with multiple off-center redshifts but no measured central redshift, we use the weighted average of the off-center redshifts as a proxy for the central redshift and then derive the redshift difference. Scaling the difference in redshift by the speed of light gives an initial estimate of a galaxy's rotational velocity, $V^0_{\mathrm{rot}}$. If the galaxy's axis of rotation, however, is not perpendicular to our line of sight, we will only measure a portion of the galaxy's total rotational velocity, namely, the portion  projected along our line of sight. To correct for this, one can normalize the rotational velocity according to its inclination, $i$, by dividing the rotational velocity by sin($i$) where $i$ is defined via the equation
\begin{equation}
    \mathrm{cos}(i) = \sqrt{\frac{(b/a)^2 - q^2}{1 - q^2}}
\end{equation}
where $q=0.2$ is chosen to be the smallest discernible axis ratio in conformity with that of the TF calibration. We therefore arrive at a measurement of the TF observable, $V_{\mathrm{rot}} = V^0_{{\mathrm{rot}}} / \mathrm{sin}(i)$.

The second observable -- the absolute $r$-band magnitude, $M$ -- can be derived from $z_{\mathrm{cmb}}$ and $m_r$, the galaxy's measured apparent magnitude (corrected for Milky Way extinction). These three quantities are related via the following equation
\begin{equation}\label{M_obs}
\begin{aligned}    
    M &= m_r - k_r - A_{\mathrm{dust}} - 5 \cdot \mathrm{log}_{10}\left[\frac{d_L(z_{\mathrm{cmb}}, h)}{10 \: \mathrm{pc}}\right]\\
    &= m_r - k_r - A_{\mathrm{dust}} - 5 \cdot \mathrm{log}_{10}\left[\int^{z_{\mathrm{cmb}}}_0 \frac{10^3 c \mathrm{d}z'}{E(z') \: [\mathrm{km~s^{-1}}]}\right] + 5 \cdot \mathrm{log}_{10}(h)\\
\end{aligned}
\end{equation}
where $d_L(z_{\mathrm{cmb}}, h)$ is a function for computing the luminosity distance, $E(z')$ is the dimensionless Hubble parameter, $k_r$ is a k-correction factor to the apparent magnitude computed using the \texttt{kcorrect} package \citep{kcorrect}, $A_{\mathrm{dust}}$ is the DR1 TFR internal dust correction dependent on axis ratio, and $h$ is defined in terms of the Hubble constant as $h = \frac{H_0}{100} ~\mathrm{km~ s^{-1}~Mpc^{-1}}$. At this point in the analysis, we assume $h=1$, but this assumption is later undone during the zero-pointing process (see Section \ref{subsection:zero_point}). $M$ directly depends upon the total redshift, $z_{\mathrm{cmb}}$, and therefore is perturbed from the true absolute magnitude to a degree proportionate to its peculiar redshift. The two TF observables are related via the following calibration equation
\begin{equation}\label{eq:M_tf_cal}
    M_{\mathrm{cal}} = a \cdot \mathrm{log}_{10}\left(\frac{V_{\mathrm{rot}}}{V_0}\right) + b
\end{equation}
where $a=-7.16 \pm 0.01$ is the slope parameter and $b=-20.59 \pm 0.01$ is the y-intercept derived using the DESI DR1 PV calibration sample while $V_0$ is a constant chosen to be $\sim150$ km s$^{-1}$, which is the median rotational velocity of calibration sample. The DR1 TF calibration fits a different y-intercept, $b$, for different redshift regimes; following the DR1 analysis we assume a value for $b$ from the lowest DR1 redshift regime ($0.03<z_{\mathrm{cmb}}<0.035$) even though the redshifts of the current analysis are smaller than 0.03. For a given galaxy, the calibrated $M_{\mathrm{cal}}$ is assumed to capture the galaxy's true absolute magnitude, which, unlike $M$, is unperturbed by the galaxy's peculiar velocity. Using the definition of log-distance ratio in Equation \ref{eq:log_distance} as well as the conversions between luminosity distance, distance modulus, and absolute magnitude, one can arrive at the following equation
\begin{equation}\label{eq:tf_ldr}
\eta_{\mathrm{gal},\mathrm{TF}} = \frac{M_{\mathrm{cal}} - M}{5}
\end{equation}
where $\eta_{\mathrm{gal},\mathrm{TF}}$ is the log-distance ratio of a single spiral galaxy in our TF sample. These log-distance ratios can then be used for the standard siren analysis.

As was the case for our FP log-distance ratios, our TF sample is not significantly affected by a selection bias that stems from our apparent magnitude cut of 18. We come to this conclusion by using a conservative estimate for the maximum group galaxy distance based on the GW distance -- $d_{L, \mathrm{GW}} + 5 \cdot \sigma_{d_{L, \mathrm{GW}}} = 49.2$ Mpc \citep{palmese_ag} -- to calculate a maximum observable absolute magnitude. The DESI DR1 TF relation is  calibrated for a population of galaxies with absolute magnitudes below $-17$. Based on our apparent magnitude threshold and maximum group galaxy distance, we expect to observe all group galaxies below an absolute magnitude of $-16$. We therefore observe all relevant TF group galaxies such that no Malmquist bias correction is required.

We use Monte Carlo resampling to derive an uncertainty on $\eta_{\mathrm{gal},\mathrm{TF}}$. The underlying parameters with non-negligible uncertainties in Equation \ref{eq:tf_ldr} are $m_r$, $A_{\mathrm{dust}}$, $z_{\mathrm{cmb}}$, $a$, $b$, and $V_{\mathrm{rot}}$. Additionally, there is an intrinsic scatter, $\sigma_{\mathrm{TF}} = 0.44$, on the TF relation that contributes to the uncertainty on $M_{\rm cal}$. We take $10,000$ samples of each of these underlying parameters from their respective Gaussian posterior distributions. From these posterior samples, we compute a posterior distribution for $\eta_{\mathrm{gal},\mathrm{TF}}$ with the width of a skew-normal fit to this distribution serving as uncertainty on $\eta_{\mathrm{gal},\mathrm{TF}}$.

\subsection{Log-Distance Ratio Zero Point}\label{subsection:zero_point}
The FP and TF relations both assume that the observables derived from the respective relations are not significantly perturbed by peculiar velocity. This assumption can be reformulated as the following claim: the average log-distance ratio of each calibration sample is negligible. Due also to the fact that DESI is not an all-sky survey, this claim is unlikely to be true for either the FP or TF calibration sample. One could, however, make this claim true by adding a constant offset to our measured log-distance ratios such that the overall average log-distance ratio of each calibration set is zero. There are a number of methods for determining such an offset with each being accompanied by its own set of assumptions and potential biases. Following the DR1 analysis outlined in \citet{DESI_zero_point}, we zero-point our log-distance ratios using the Type Ia supernovae in \citet{Riess_2022} and \citet{Pantheon+}. This procedure amounts to an undoing of the assumed fiducial cosmology in the FP and TF analyses in which $H_0=100$ km s$^{-1}$ Mpc$^{-1}$. The DR1 analysis derived that all FP log-distance ratios ought to be shifted down $\Delta \eta_{\mathrm{zp}} = 0.139 \pm 0.007$, and we apply this offset to our FP log-distance ratios accordingly.

For the TF galaxies, we derive a zero-pointing value that departs slightly from that derived in the DR1 analysis, ultimately leading to our choice to exclude these galaxies from our fiducial Hubble constant analysis. The DR1 analysis applies a shift to TF log-distance ratios that enforces agreement between the average TF and FP log-distance ratios of galaxies in the same group. The justification for this shift is that one would expect the average log-distance ratio calculated from TF galaxies in a given group to be the same as that calculated from FP galaxies in the same group. In the DR1 analysis, the TF log-distance ratios are shifted by $\Delta \eta_{\mathrm{TF}} = -0.005 \pm 0.004$. In NGC 4993's group, after applying $\Delta \eta_{\mathrm{zp}}$, we observe a much larger discrepancy than the DR1 $\Delta \eta_{\mathrm{TF}}$ between the weighted average TF log-distance ratio ($-0.11 \pm 0.04$) and the weighted average FP log-distance ratio ($0.11 \pm 0.04$). The large discrepancy is in line with the TF-FP divergence in the DR1 data at the redshifts of these galaxies ($0.0095 < z_{\mathrm{cmb}} < 0.0125$). These redshifts are towards the edge of the fiducial DR1 redshift regime ($z_{\mathrm{cmb}}=0.01$) where \citet{DESI_zero_point} report known issues aligning log-distance ratios from DESI and other calibrators (such as supernovae). We calculate the weighted average of the TF and the FP galaxies in the $0.0095 < z_{\mathrm{cmb}} < 0.0125$ regime and calculate the discrepancy between them to be $\Delta \eta_{\mathrm{TF}} = -0.05 \pm 0.02$. This discrepancy is then applied as a shift to all TF log-distance ratios. We note that our $H_0$ measurements neglect the systematic uncertainties as they are all at least an order of magnitude smaller than our average statistical uncertainties.

We choose to apply the $\Delta \eta_{\mathrm{TF}}$ shift to the TF log-distance ratios rather than FP log-distance ratios %for a number of reasons that impugn the reliability of the TF measurements in the NGC 4993 group. First, there are two FP galaxies and two TF galaxies that are measured in both DESI and Cosmicflows-4 (CF4) as shown (with $\Delta \eta_{\mathrm{TF}}$ applied to the DESI TF measurements) in Figure~\ref{fig:{phase-space}}. Without the $\Delta \eta_{\mathrm{TF}}$ correction, both DESI TF measurements are discrepant with CF4 at the $1\sigma$ level while the FP galaxies show good agreement between DESI and CF4. Although the DESI-CF4 TF disagreement could very well be due to statistical measurement fluctuation, this disagreement suggests that the TF galaxies might be a better candidate for correction than the FP galaxies. We acknowledge, however, that this suggestion relies on the trustworthiness of CF4 measurements as an independent check on our DESI measurements. An additional reason to apply $\Delta \eta_{\mathrm{TF}}$ to TF galaxies rather than FP galaxies relates to
because of the redshift regime of the DESI TF calibration. As mentioned in Section \ref{subsection:fp_method}, the calibrated TF y-intercept derives from data at a redshift which is at least 0.03, which is higher than the redshifts in the NGC 4993 group. In fact, the final DR1 TF sample includes very stringent rotational velocity and apparent magnitude cuts that no galaxies at NGC 4993's redshift are included in the DR1 sample. For this reason, we surmise that the FP calibration is more reliable at the NGC 4993 group redshift and choose to apply the $\Delta \eta_{\mathrm{TF}}$ shift to our TF log-distance ratios.
% See Figure 13 in DR1 TF paper that shows our redshift regime is excluded.

Even after adding the $\Delta \eta_{\mathrm{TF}}$ shift, the discrepancy between the TF and FP galaxies in the NGC 4993 group remains suspiciously large. To quantify this effect, we perform a Mann-Whitney U analysis \citep{mann1947test} to determine if the FP and shifted TF peculiar velocities derive from the same underlying distribution. We use each peculiar velocity measurement and uncertainty to construct a normal distribution from which we select one sample. We use the \texttt{mannwhitneyu} function from \texttt{scipy} to extract a $p$-value associated with the null hypothesis that these two datasets derive from the same distribution. We perform this procedure 10,000 times leading to 10,000 $p$-values, with the scatter in these values deriving from the peculiar velocity measurement uncertainties. The $1\sigma$ range of $p$-values spans 0.03 to 0.34, indicating that if the two datasets derived from the same distribution, we would expect to see a discrepancy this large 3\% to 34\% of the time. The Mann-Whitney U analysis therefore shows it is more likely  that either the TF or the FP galaxies might not trace the underlying peculiar velocity distribution of the NGC 4993 galaxy group. Although the discrepancy may still be attributed to a statistical fluctuation, in light of the fact that the TF peculiar velocities cannot be trusted at the redshifts of interest (as shown for the DESI DR1 analyses), we choose to not include the TF log-distance ratios as part of our fiducial $H_0$ analysis. Instead, we let our fiducial dataset only consist of FP measurements and compare with the $H_0$ result derived from only TF measurements.

\section{Results}\label{section:results}

\begin{figure*}[htbp]
  \centering
    \includegraphics[width=\linewidth]{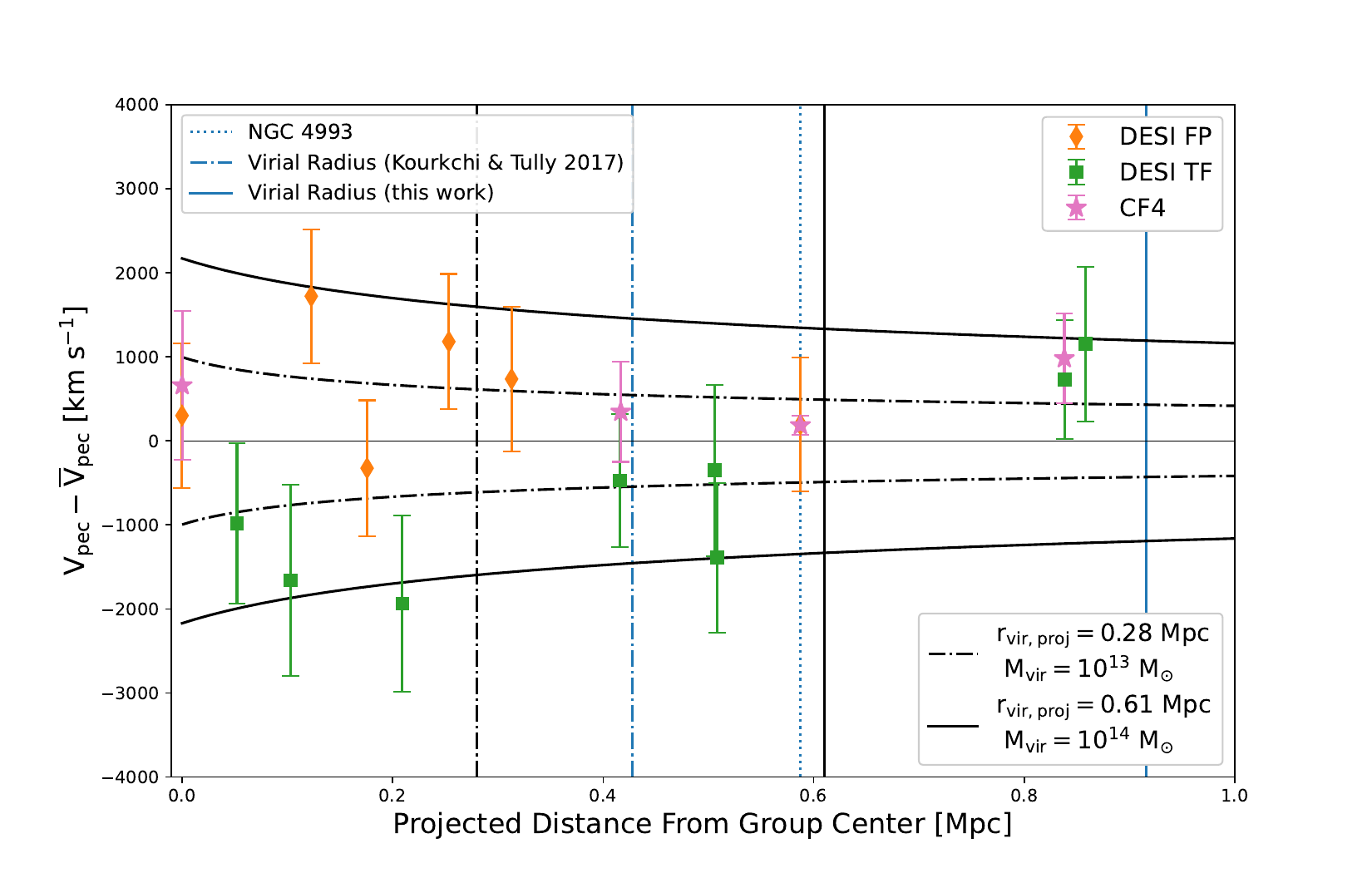}
    \caption{\textbf{Peculiar velocity phase-space diagram of group galaxies.} The x-axis displays each galaxy's projected distance from the brightest group galaxy (an approximation of the group center) where all galaxies are assumed to be at the same comoving distance. The y-axis indicates each galaxy's peculiar velocity offset by the weighted average peculiar velocity. We also display peculiar velocities from group galaxies in CF4 for comparison. We demarcate escape velocity envelopes for galaxy groups of differing halo masses along with their corresponding projected virial radii. Additionally, we delineate the projected virial radii measured from galaxy samples in \citet{kourkchi_ngc4993_pv} and this work.}\label{fig:{phase-space}}
\end{figure*}

\subsection{Observed Peculiar Velocity}\label{subsection:pv_res}
The original DESI dataset of observed PV targets in the group includes nine FP galaxies and 17 TF galaxies. We remove three FP galaxies from our sample since they possess velocity dispersions well below the DESI spectrograph resolution limit of 50 km s$^{-1}$. We also remove nine TF galaxies from our sample since they exhibit TF-calibrated absolute magnitudes greater than the cutoff of $-17$. In \citet{douglass_tfr_edr}, all galaxies with absolute magnitudes above the cutoff are considered dwarf galaxies for which the calibrated TF relation does not hold. In order to preserve the purity of the calibration cutoff, \citet{douglass_tfr_Y1} make this cutoff more stringent, requiring that all galaxy apparent magnitudes be larger than the minimum of 17.75 and $\mu(z_{\mathrm{cmb}}, h) - 17 + 5 \: \mathrm{log_{10}} h$ where $\mu(z_{\mathrm{cmb}}, h)$ is the distance modulus at redshift $z_{\mathrm{cmb}}$. Because our analysis does not require the same level of purity as the calibration sample, we adopt the earlier cutoff value, thereby permitting a larger number of galaxies to be included in our sample. After these quality cuts, the fiducial sample consists of six FP galaxies and eight TF galaxies. At this stage, we convert the log-distance ratios in this fiducial sample into peculiar velocities using Equation \ref{eta_to_pv} in order to plot the phase space diagram shown in Figure~\ref{fig:{phase-space}}.

We layer Figure~\ref{fig:{phase-space}} with group escape velocity curves based on a Navarro–Frenk–White profile (assuming $h=0.7$) given various dark matter halo virial radii (and corresponding dark matter halo virial masses). We also calculate a projected virial radius, $r_{\mathrm{vir}, \mathrm{proj}} \approx 0.92$ Mpc, from our observed galaxy dataset using the equation
\begin{equation}\label{eq:vir_proj}
    r_{\mathrm{vir}, \mathrm{proj}} \approx \frac{N^2}{\sum_{i>j} 1 / R_{ij}}
\end{equation}
where $N=39$ is the number of galaxies observed in the group and $R_{ij}$ is the projected distance between the $i$th and $j$th galaxies. In Figure~\ref{fig:{phase-space}}, we plot our value of $r_{\mathrm{vir}, \mathrm{proj}}$ along with the projected virial radius from \citep{kourkchi_ngc4993_pv}. We note that most characterizations of the virial radius (except for our own) would exclude two TF galaxies as group members. We also include the peculiar velocity measurements from CF4 for four galaxies contained in both our DESI sample and CF4. We note that all peculiar velocities in this plot are offset so that the weighted mean peculiar velocity of DESI galaxies in the plot is zero.

Before measuring the observed log-distance ratio, $\eta_{\mathrm{obs}}$, to be used in the standard siren $H_0$ posterior, we construct a number of group galaxy samples in addition to the fiducial sample, which includes the log-distance ratios of all DESI FP galaxies plotted in Figure~\ref{fig:{phase-space}}. We measure $H_0$ for each distinct sample in order to understand the effect of different choices on our result. The log-distance ratios and peculiar velocities of these sub-samples are tabulated in Table \ref{tab:results_samples}; shorthand designations for each sub-sample are shown in parentheses. We first construct a sub-sample consisting of only TF galaxies (``Tully-Fisher''). We then construct a subset that conforms to the group redshift regime claimed in \citep{hjorth_2017} to examine if our more lax redshift cuts affect the $H_0$ posterior (``H17''). Our next subset consists of galaxies that lie within the projected virial radius of \citet{kourkchi_ngc4993_pv} (``KT17 $r_{\mathrm{vir},\mathrm{proj}}$''). The final subset invokes the following three ``strict'' criteria (``Strict''). In this sample, we exclude one FP galaxy that narrowly fails a calibration sample cut requiring the difference between the $g$-band and $r$-band absolute magnitude to be greater than 0.68 \citep{target_selection}. Given that the determination of quality cut bounds requires some discretion judgment, it is reasonable to apply the FP calibration to this one galaxy since it conforms to all other FP magnitude and velocity dispersion quality controls. Yet, we exclude this galaxy from our strict sample. By measuring $H_0$ from these different samples, we can discern potential biases relating to inclusion and exclusion of certain galaxies from the group.

For each of our samples, there remains a question of how exactly one ought to combine the log-distance ratios of many galaxies into a single value for $\eta_{\mathrm{obs}}$. In previous studies, a smoothing kernel centered on the sky location of NGC 4993 was incorporated into the calculation of $\eta_{\mathrm{obs}}$ in order to upweight the log-distance ratios (or, equivalently, the peculiar velocities) of galaxies closer to NGC 4993 \citep{gw170817_nature, nicolaou, standard_siren_speeds}. For some of these studies, an entire catalog of galaxies was included in the determination of $\eta_{\mathrm{obs}}$, so the incorporation of a smoothing kernel ensured that galaxies not endemic to the group did not significantly affect the $H_0$ measurement. We eschew the smoothing kernel in our work in favor of determining $\eta_{\mathrm{obs}}$ via a simple weighted average of the measured log-distance ratios. Since $\eta_{\mathrm{obs}}$ is meant to be a measure of the \textit{group's} log-distance ratio, no preferential weighting should be given to galaxies in the NGC 4993 locale. Furthermore, so long as the galaxies we include in our weighted average are actually group galaxies, there is no reason to penalize galaxies farther away from NGC 4993.

We also construct two additional samples from which we compute $\eta_{\mathrm{obs}}$ based on observations that are external to DESI. One measurement computes $\eta_{\mathrm{obs}}$ from the Cosmicflows-3 (CF3) distance moduli of three group galaxies \citep{cf3} that were designated as such in \citep{tully_ngc4993_pv}. We use CF3 rather than CF4 in order to compare to \citet{standard_siren_speeds}, which predates CF4. The second measurement derives from the distance modulus of NGC 4993 itself -- rescaled slightly in CF4 \citep{cf4} to minimize disagreement with other distance indicators -- as measured via Surface Brightness Fluctuations (SBF) \citep{sbf_ngc4993}. Although \citet{sbf_ngc4993} used this SBF distance to measure an $H_0$ value, our work is the first to utilize SBF to measure the peculiar velocity of NGC 4993 in the context of a standard siren $H_0$ measurement. In both the CF3 and SBF samples, we convert distance modulus to a log-distance ratio using the $H_0$ measured in from overall Cosmicflows catalog, either CF3 or CF4. These two non-DESI samples provide an important point of comparison for our ultimate $H_0$ posterior results.

\subsection{Group Membership and Virialization}\label{subsection:cluster_mem}

After analyzing the dynamics of galaxies near NGC 4993, it becomes natural to ask: Is NGC 4993 actually in a galaxy group? Previous studies have noted that NGC 4993 shows signs of having recently undergone a galaxy-galaxy merger \citep{Palmese_2017,4993_galaxy_merger}, which are more frequent in galaxy groups compared to isolated galaxies, potentially supporting the point that NGC 4993 is indeed in a larger structure. As noted in \citet{bin_host_con}, NGC 4993 may lie at the outskirts of the group when the center is taken to be the brightest group galaxy. This is surprising in light of the relatively high stellar mass of NGC 4993 ($(3.8 \pm 0.20)\times 10^{10}~M_\odot$; \citealt{Palmese_2017}), as dynamical friction would facilitate the infalling of massive galaxies towards the center.  We further explore this question of group membership in this subsection.

Until this juncture, we have only invoked two criteria to determine whether or not a galaxy is in the same group as NGC 4993. We first require that the galaxy is a part of our observations, meaning it lies within $\sim$2.3 degrees of NGC 4993 on the sky. Secondly, we require that the galaxy has a redshift between 0.008 and 0.012. Using these criteria, we observe 39 galaxies -- of which 14 are peculiar velocity targets -- that ought to be considered as part of the group. The first of our criteria is relatively imprecise since it is set by the size of the DESI tiles used to take the observation. The second, however, is justified by the redshift-space diagram of our observations in Figure~\ref{fig:{redshift-space}}. Previous work \citep{hjorth_2017} had chosen more stringent bounds, roughly equivalent to (0.00934, 0.01002), based upon redshift-space gaps that had been observed beyond these bounds. In our redshift-space diagram, no clear gap appears until our chosen boundaries are reached, justifying our choice of redshift regime.

Yet, satisfaction of these initial criteria is not sufficient to justify a galaxy's membership in the galaxy group. In search of such justification, we appeal to Figure~\ref{fig:{phase-space}}. If a galaxy is truly part of the group, then it should be gravitationally-bound to the group. In terms of velocity, we expect the galaxy to possess a peculiar velocity in the group rest frame smaller than the maximum escape velocity predicted from the group's dark matter halo mass. The peculiar velocities in Figure~\ref{fig:{phase-space}} are approximately in the rest frame of the group since the weighted mean group velocity has been subtracted off. Thus, the remaining peculiar velocity components displayed in Figure~\ref{fig:{phase-space}} are peculiar to the individual galaxies rather than the group as a whole. The galaxies in Figure~\ref{fig:{phase-space}} seem to trace out a galaxy group with a suggested halo mass of order $10^{13}-10^{14} M_\odot$ with two galaxies falling well beyond the projected virial radius corresponding to a halo mass of order $10^{14} M_\odot$. While some previous work has suggested a halo mass closer to $10^{12} M_\odot$ \citep{bin_host_con}, \citet{kourkchi_ngc4993_pv} report a virial radius that would be more in line with a halo mass between $10^{13} M_\odot$ and $10^{14} M_\odot$. %We also emphasize that the group galaxy catalog in \citep{kourkchi_ngc4993_pv} displays a number of groups with similar halo mass for groups with approximately 40 galaxies (and even less). 
We note that if only FP galaxies are considered, as done in the following sections for the fiducial standard siren analysis, the scatter is slightly reduced but the group remains consistent with a $\sim 10^{13} M_\odot$ halo. A similar conclusion is reached when analyzing the scatter of the approximated recessional velocities $cz$ of all potential group galaxies around their mean value. Although Figure~\ref{fig:{phase-space}} is by no means conclusive, it conforms to the claim that our measured peculiar velocities derive from group galaxies and that the halo mass is of order $10^{13} M_\odot$.

Next, we consider the specific position of NGC 4993 in the group. Based on the velocity phase-space diagrams we investigate, if NGC 4993 were to be the central of the group halo, it would exclude a number of candidate group  members, which are otherwise consistent with being part of the group when the brightest galaxy is instead taken as the center. In our fiducial analysis, NGC 4993 is therefore placed in the outskirts of the galaxy group. Given also the $\sim 10^{13}~M_\odot$ halo mass and the relatively high mass of NGC 4993, it is possible that NGC 4993 has only recently fallen into the group and experienced a galaxy merger \citep{Palmese_2017} as part of the process. 

One might still question whether these supposed ``group galaxies'' even form a group with peculiar velocities scattered about a coherent group peculiar velocity component. To determine whether the peculiar velocities of these galaxies are coherent, one can examine whether the collection of galaxies is virialized. To probe the hypothesis that this group is relaxed, we perform a Dressler-Shectman (DS) test \citep{ds_dressler, ds_halliday} as an examination of potential sub-structure. Following \citet{tests_of_virialization}, we calculate the DS statistic, $\Delta$, using the equation
\begin{equation}
    \Delta = \sum^{39}_{i=1} \frac{11}{\sigma^2_{\mathrm{gr}}}[(\overline{v}_i - \overline{v}_{\mathrm{gr}})^2 + (\sigma_i - \sigma_{\mathrm{gr}})^2]
\end{equation}
where $\overline{v}_i$ and $\sigma_i$ are the average and standard deviation of the total observed line-of-sight velocities from the 10 galaxies in closest proximity to $i$th galaxy in the collection with the index $i$ running over all 39 unique galaxies in our collection. $\overline{v}_{\mathrm{gr}}$ and $\sigma_{\mathrm{gr}}$ are the average and standard deviation of all the velocities in the collection. To display the significance of $\Delta$, we randomly assign each of the 39 velocities to each of the 39 galaxy coordinates and compute $\Delta$, repeating this procedure 2000 times. Using these randomly sampled $\Delta$ values, we compute a percentile in which the $\Delta$ of the actual galaxy collections falls. We find $\Delta = 48.53$ where $90.0\%$ of random runs produce a $\Delta$ below our measured value. Since a higher $\Delta$ indicates less sub-structure, our DS test suggests that there is little evidence of sub-structure in the galaxy collection. If there was evidence of sub-structure, it would suggest that the collection was not relaxed. We emphasize that even without the presence of sub-structure, the collection could still not be relaxed.

The relaxation hypothesis can also be probed by estimating the relaxation time for collection of galaxies given their characteristic radius and velocity dispersion. One can make a rough estimation of the relaxation time, $t_{\mathrm{relax}}$, using the equation
\begin{equation}
    t_{\mathrm{relax}} \approx \frac{R_{\mathrm{vir}}}{\sigma_{\mathrm{gr}}} \cdot \frac{N}{8\mathrm{ln}(N)}
\end{equation}
where $R_{\mathrm{vir}}$ is the virial radius of the group and $N$ is the number of galaxies in the group. Previous work has ostensibly used this approximation to estimate a large relaxation time ($10^{10}$ years), suggesting that the collection of galaxies might not be relaxed \citep{hjorth_2017}. Using our observations which incorporate more galaxies, we calculate a similar estimate using our measured values of $R_{\mathrm{vir}} \approx \frac{\pi}{2} \cdot r_{\mathrm{vir}, \mathrm{proj}} \sim 1.4$ Mpc,  $\sigma_{\mathrm{gr}}\sim230$ km s$^{-1}$, and $N=39$; this results in a relaxation time of $\sim8\cdot10^9$ years. For the purposes of comparison, we also examine the case where the group is in reality very small with $N=5$ members, $R_{\mathrm{vir}} \sim 0.26$ Mpc, and $\sigma_{\mathrm{gr}}\sim190$ km s$^{-1}$. This characterization results in a relaxation time of $\sim5\cdot10^8$ years. Given that the age of the universe is estimated to be $1.38 \cdot 10^{10}$ years old \citep{planck18}, it is possible that the collection of galaxies in question has reached a stage of relaxation (and even more likely if the group is smaller than we have observed). Like the DS test, however, this approximation leaves the question of relaxation unresolved.

In an attempt to reach a conclusion regarding the question of this group's relaxation and virialization, we assemble a number of pieces of evidence -- a redshift-space diagram in Figure~\ref{fig:{redshift-space}}, a phase-space diagram in Figure~\ref{fig:{phase-space}}, a Dressler-Shectman test, and an approximation of the relaxation time. Based on this set of evidence, we cannot resoundingly claim that this group is or is not virialized.

\begin{table}[htbp]
      \centering
        \begin{tabular}{l|ccc}
        \hline
        & $\eta_{\mathrm{obs}}$ & $v_{\mathrm{pec}}$ & $H_0$ \\
        & & (km s$^{-1}$) & (km s$^{-1}$ Mpc$^{-1}$) \\
        \hline
        Fiducial (FP) & $0.10^{+0.05}_{-0.04}$ & $850^{+380}_{-320}$ & $70.9^{+6.4}_{-8.5}$ \\
        No Afterglow & $0.10^{+0.05}_{-0.04}$ & $850^{+380}_{-340}$ & $72^{+14}_{-11}$ \\
        Tully-Fisher & $-0.04^{+0.05}_{-0.03}$ & $-330^{+380}_{-280}$ & $98.9^{+9.5}_{-9.9}$ \\
        H17 & $0.11^{+0.05}_{-0.05}$ & $910^{+400}_{-420}$ & $69.9^{+8.6}_{-8.0}$ \\
        KT17 $r_{\mathrm{vir},\mathrm{proj}}$ & $0.11^{+0.05}_{-0.04}$ & $930^{+380}_{-390}$ & $69.7^{+7.9}_{-8.3}$ \\
        Strict & $0.10^{+0.04}_{-0.05}$ & $800^{+340}_{-430}$ & $71.8^{+9.3}_{-6.8}$ \\
        \hline
        Cosmicflows-3 & $0.11^{+0.03}_{-0.04}$ & $930^{+270}_{-330}$ & $69.4^{+7.0}_{-6.1}$ \\
        SBF & $0.05^{+0.01}_{-0.02}$ & $360^{+110}_{-110}$ & $73.4^{+3.3}_{-3.9}$ \\
        \hline
        \citet{gw170817_nature} & -- & $310^{+150}_{-150}$ & $74^{+8}_{-12}$ \\
        \citet{nicolaou} & -- & -- & $69^{+14}_{-9}$ \\
        \citet{standard_siren_speeds} & $0.11^{+0.04}_{-0.04}$ & $860^{+300}_{-300}$ & $58.0^{+6.1}_{-5.3}$ \\
        \citet{mukherjee_2021} & -- & $390^{+130}_{-130}$ & $68.3^{+4.6}_{-4.5}$ \\
        \citet{palmese_ag} & -- & $290^{+160}_{-160}$ & $75.5^{+5.3}_{-5.4}$ \\
        \hline
        \end{tabular}
        \caption{\textbf{Results from different log-distance ratio samples.} We provide the standard siren posterior values for $\eta_{\mathrm{obs}}$, log-distance ratio; $v_{\mathrm{pec}}$, peculiar velocity; and $H_0$. We delineate three types of posteriors: 1) posteriors from different galaxy sub-samples of our DESI data (see Section \ref{subsection:pv_res} for the criteria of each sub-sample) 2) results from our own posterior calculations of two non-DESI samples – CF3 group galaxies and a single SBF measurement of NGC 4993 from CF4 3) posterior values (when available) from prior studies. Note that the results from \citet{nicolaou} and \citet{palmese_ag} are obtained after marginalization over a range of kernel sizes centered on NGC 4993.  When reporting \citet{standard_siren_speeds}, we select the results from their \citet{tully_ngc4993_pv} group, which utilizes the maximal number of reliable galaxy PVs. All reported uncertainties are at the 68\% credible intervals on the respective posteriors.}
        \label{tab:results_samples}
\end{table}

\begin{figure*}[htbp]
  \centering
    \includegraphics[width=\linewidth]{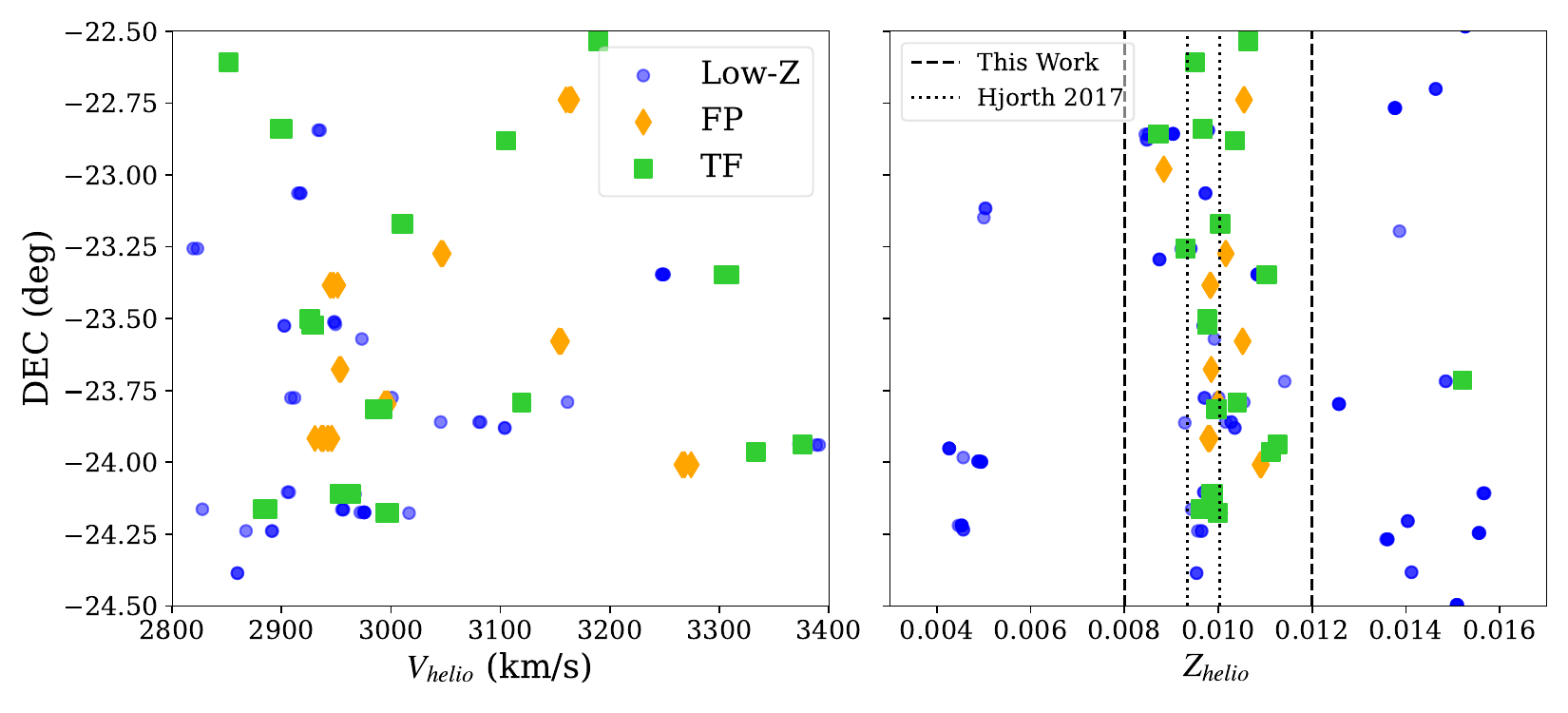}
    \caption{\textbf{Velocity- and redshift-space of observations.} Left: For each galaxy, we plot its declination versus its heliocentric velocity (i.e. the speed of light multiplied by the heliocentric redshift), distinguishing our TF (green squares) and FP (yellow diamonds) targets from the DESI Low-Redshift Survey targets (blue circles). Previous work \citep{hjorth_2017} had observed a gap in redshifts from $3005$ km/s to $3169$ km/s. Our observations demonstrate that this gap does not exist. Right: A comparison between the chosen redshift bounds of our work (dashed) and previous work \citep{hjorth_2017} (dotted). Given our observations, we select boundary edges at the points where a gap first appear in redshift space.}\label{fig:{redshift-space}}
\end{figure*}

\subsection{Standard Siren Posterior}\label{subsection:H0}
In Figure~\ref{fig:{H0}}, we plot the posterior of the Hubble constant using $\eta_{\mathrm{obs}}$ from the fiducial sample of galaxies along with the sub-samples delineated in Section \ref{subsection:pv_res}. We also display a summary of these results in Table \ref{tab:results_samples}. The fiducial constraint on the Hubble constant from FP galaxies alone is $H_0 = 70.9^{+6.4}_{-8.5}$ km s$^{-1}$ Mpc$^{-1}$, which agrees with the $H_0$ measurements from both \emph{Planck} and SH0ES at the $1\sigma$ level. Using only TF galaxies, we arrive at a constraint (not pictured in Figure~\ref{fig:{H0}}) of $H_0 =98.9^{+9.5}_{-9.9}$ km s$^{-1}$ Mpc$^{-1}$, which markedly departs from our fiducial $H_0$ measurement from FP galaxies. The $H_0$ discrepancy between these samples is attributable to their difference in log-distance ratio: The posterior group log-distance ratio (peculiar velocity) is $\eta_{\mathrm{obs}} = 0.10^{+0.05}_{-0.04}$ ($v_{\mathrm{pec}} = 850^{+380}_{-320}$ km s$^{-1}$) for FP galaxies and $\eta_{\mathrm{obs}} = -0.04^{+0.05}_{-0.03}$ ($v_{\mathrm{pec}} = -330^{+380}_{-280}$ km s$^{-1}$) for the TF galaxies.

Comparing our fiducial measurement to the other peculiar velocity sub-samples we measured, we find minimal change in the location of the $H_0$ posterior. The primary difference between our fiducial and alternative sub-samples is that the latter yields broader $H_0$ constraints (due to a less constrained log-distance ratio) with little evidence of a systematic shift in the posterior value. Still, one can observe that even these small variations in $\eta_{\mathrm{obs}}$ affect $H_0$ such that $H_0$ tends to increase with decreasing $\eta_{\mathrm{obs}}$. Since including or excluding galaxies from our fiducial sample does not alter the average observed log-distance ratio, our $H_0$ posterior is quite robust to alternative sub-samples.

In addition to examining the effect on $H_0$ of using different log-distance ratio sub-samples, we also demonstrate how the different characterizations of the $H_0$ posterior calculation (as outlined in Section \ref{subsection:posterior}) affect our $H_0$ measurement. Figure~\ref{fig:{H0_fiducials}} displays the resulting $H_0$ posterior for each characterization; the label for each characterization in Figure~\ref{fig:{H0_fiducials}} is indicated in the parentheticals below. We examine the effect on $H_0$ when 1) we incorporate a free parameter that accounts for the difference in distance between NGC 4993 and the group center (``$d_L$ Perturbation''); 2) we weight galaxy log-distance ratios using a smoothing kernel (``Smoothing Kernel''); 3) we center the group at NGC 4993 and use its redshift as the group redshift (``NGC 4993 Center''); and 4) we utilize a smoothing kernel while \textit{also} centering the group at NGC 4993 (``NGC 4993 Center \& Smoothing Kernel''). We first note that adding a smoothing kernel -- either in our fiducial or NGC-4993-center case -- does not drastically alter the $H_0$ posterior. The result is unsurprising given the relatively small scatter between FP PVs shown in Figure~\ref{fig:{phase-space}}. Similarly, we find that the posterior that uses the ``$d_L$ Perturbation'' method differs very little from the fiducial $H_0$ posterior. One may have expected to observe a broadening of the posterior upon adding the free perturbative term since it is degenerate with most of the other posterior parameters. We find, however, that any broadening of the posterior from this additional parameter is negligible. In contrast, the ``NGC 4993 Center'' method distinctively moves $H_0$ to lower values. This alteration relates to the issue (described in Section \ref{subsection:H0}) of using both group and merger host quantities in the standard siren analysis, specifically the value used for $z_{\mathrm{tot}}$ in Equation \ref{eq:log_distance}. In this posterior analysis we assume $z_{\mathrm{tot}}$ to be the DESI-measured redshift of NGC 4993, $z_{\mathrm{tot}} = (9.83 \pm 0.03) \cdot 10^{-3}$, rather than our fiducial measurement of the brightest central group galaxy redshift $z_{\mathrm{tot}} = (1.090 \pm 0.003) \cdot 10^{-2}$. The difference between these redshifts outpaces the precision on either of the redshifts. If one wishes a redshift compatible with the GW luminosity distance of the merger, then one ought to use the NGC 4993 redshift if a peculiar velocity measurement of NGC 4993 alone is also available. If, however, one prefers a redshift that conforms to the group peculiar velocity used in the standard siren analysis, one ought to use the central group galaxy redshift. Since in our fiducial analysis we use the group's peculiar velocity, and since we have deemed the group center to be consistent with the brightest group galaxy, we suggest that the latter option is the appropriate one in this case. It is also important to note that the different choices are still consistent within the uncertainties. 

For the purposes of comparison to other GW170817 standard siren measurements, we also display the $H_0$ posterior without the degeneracy-breaking afterglow data in Figure~\ref{fig:{H0_prior_work}}. Expectedly, the $H_0$ constraint in this case is much broader than our fiducial constraint. In Figure~\ref{fig:{H0_prior_work}}, we also display an $H_0$ posterior using log-distance ratios from CF3 galaxies and using an SBF-measured PV of NGC 4993 alone (as described in Section \ref{subsection:pv_res}). The corresponding $H_0$ constraints from each of these samples in listed in Table~\ref{tab:results_samples}. In fact, the tightest constraint in Table~\ref{tab:results_samples} comes from the SBF-measured PV with $H_0 = 73.4^{+3.3}_{-3.9}$ km s$^{-1}$ Mpc$^{-1}$. We compare these results (along with our fiducial results) with the $H_0$ posteriors from previous work in Section \ref{section:Discussion}.

\begin{figure*}[htbp]
  \centering
    \includegraphics[width=\linewidth]{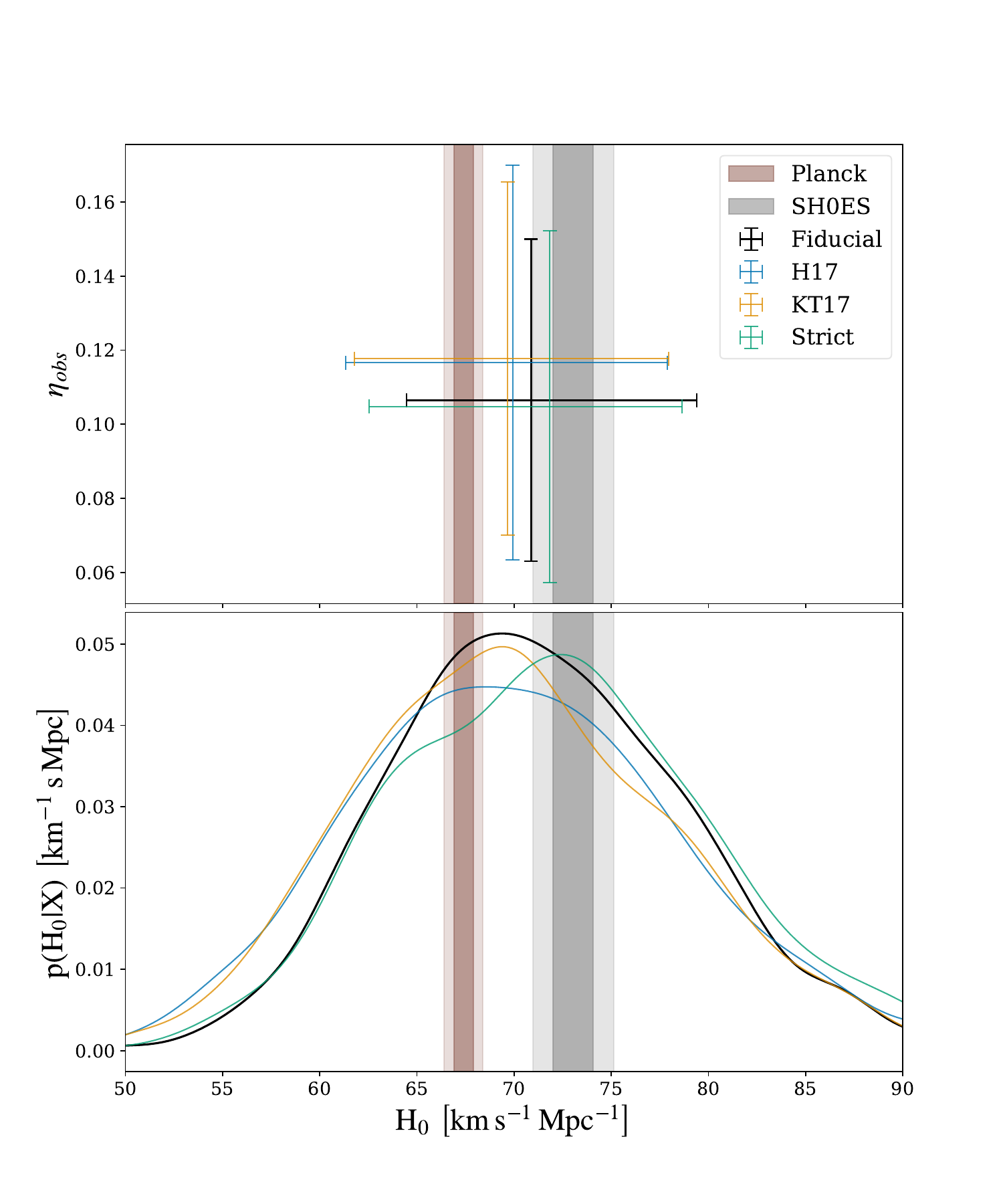}
    \caption{\textbf{$H_0$ posteriors with varying PV samples.} Lower panel: $H_0$ posteriors for a number of DESI galaxy sub-samples. See Section \ref{subsection:pv_res} for details about the selection criteria of each sub-sample. Upper panel: For each sub-sample, the posterior group-average log-distance ratio is plotted against the posterior $H_0$.}\label{fig:{H0}}
\end{figure*}

\begin{figure*}[htbp]
  \centering
    \includegraphics[width=\linewidth]{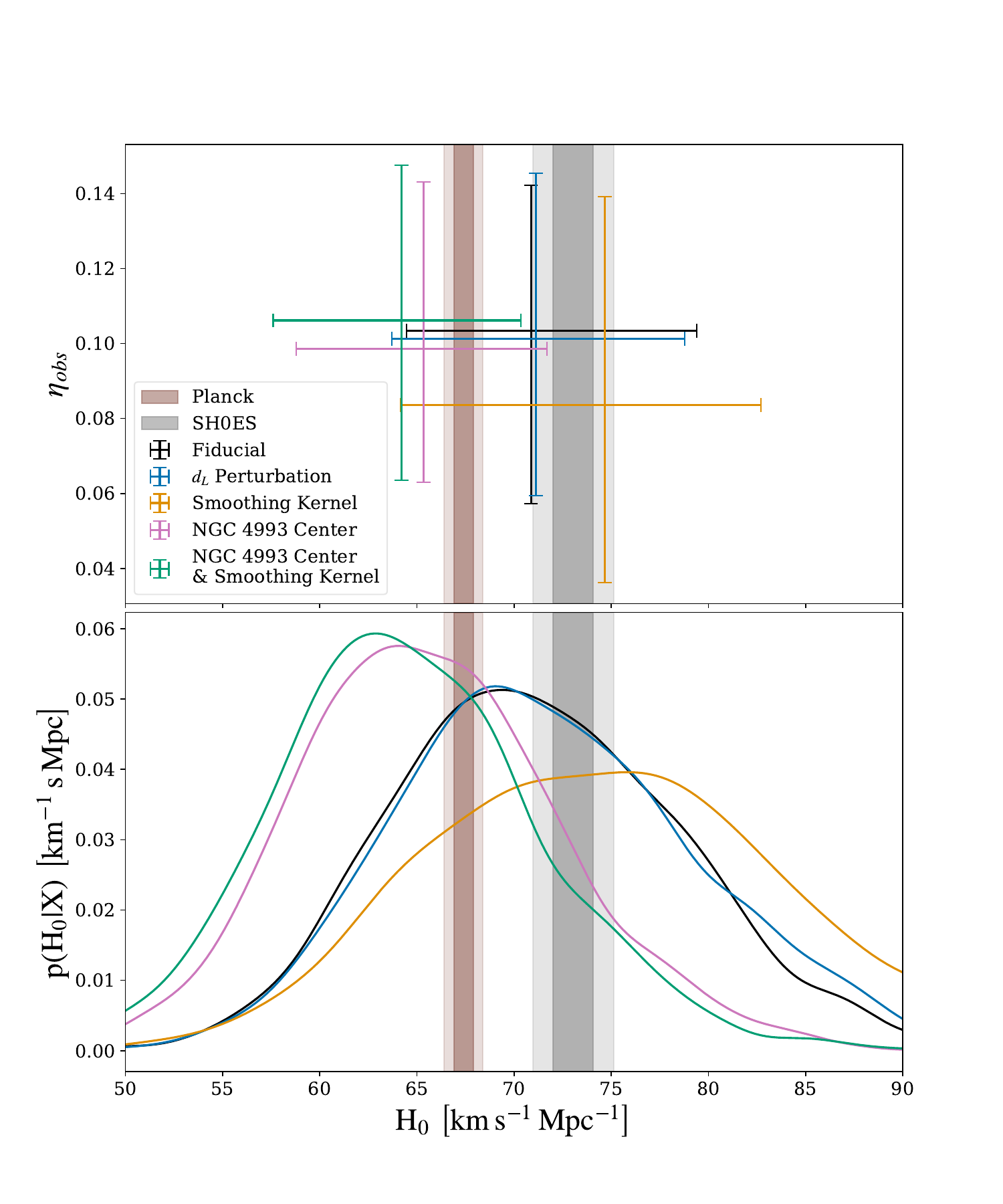}
    \caption{\textbf{$H_0$ posteriors with varying methods.} Lower and upper panels are similar to those in Figure~\ref{fig:{H0}}. The different posteriors derive from the fiducial posterior characterization (``Fiducial''), a parameterization that accounts for the comoving distance between NGC 4993 and the group center (``$d_L$ perturbation''), a posterior that incorporates a smoothing kernel into the log-distance ratio averaging (``Smoothing Kernel''), a posterior with the group centered on the coordinates of NGC 4993 (``NGC 4993 Center''), and a posterior with a smoothing kernel centered on NGC 4993 (``NGC 4993 Center \& Smoothing Kernel''). The vertical bands represent the 1 and $2\sigma$ constraints from SH0ES and \emph{Planck}, respectively.}\label{fig:{H0_fiducials}}
\end{figure*}

\begin{figure*}[htbp]
  \centering
    \includegraphics[width=\linewidth]{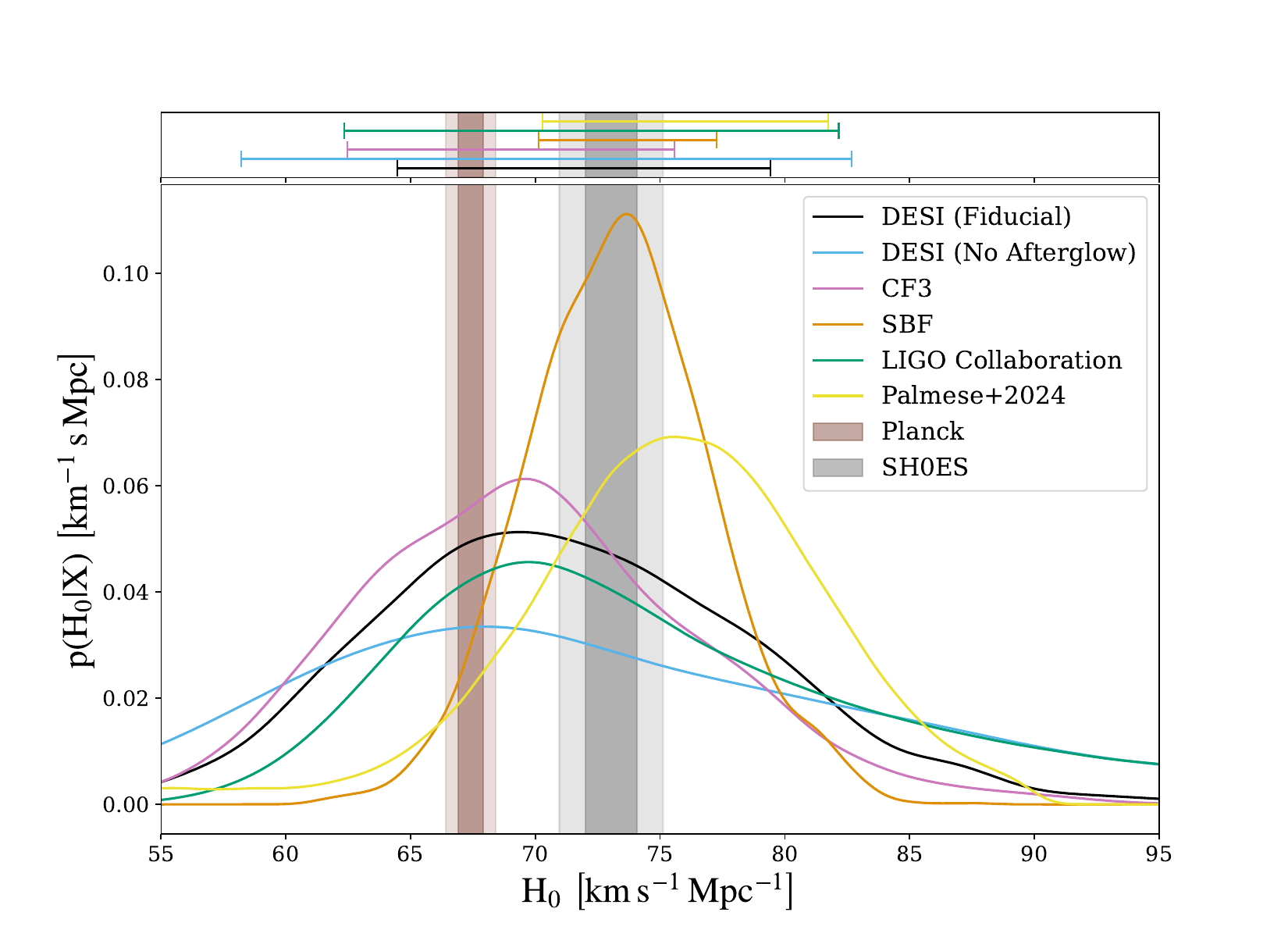}
    \caption{\textbf{$H_0$ posterior comparison with non-DESI PV samples and prior studies.} We compare our Fiducial DESI result (with and without the inclusion of $H_0$-constraining afterglow information) to the original standard siren result from \citet{gw170817_nature} and a more recent measurement from \citet{palmese_ag}. We also show posterior that we have computed using peculiar velocity samples from group galaxies in CF3 and a SBF-based PV measurement of NGC 4993 alone from CF4.}\label{fig:{H0_prior_work}}
\end{figure*}

\section{Discussion}\label{section:Discussion}
We first analyze the results from our two primary DESI samples: The fiducial FP sample and the TF sample. As Table \ref{tab:results_samples} shows, there is a $\sim2\sigma$ discrepancy between the average peculiar velocities and resulting $H_0$ measurements from these two samples. Given the issues described in Section \ref{subsection:zero_point} regarding the zero-pointing of the TF log-distance ratios, we deem the constraints from the TF sample to be unreliable compared to the fiducial constraints.

We next compare the precision of our $H_0$ constraint to previous studies. At first glance, Figure~\ref{fig:{H0_prior_work}} suggests that DESI peculiar velocities yield less precise constraints than those from other studies. Without afterglow information, the DESI result is less constrained than the original LIGO measurement, $H_0 =74^{+8}_{-12}$ km s$^{-1}$ Mpc$^{-1}$, in \citet{gw170817_nature}. Similarly, our fiducial DESI result shows broader constraints than the most recent analysis by \citet{palmese_ag}, which found $H_0 =75.5^{+5.3}_{-5.4}$ km s$^{-1}$ Mpc$^{-1}$. Furthermore, $H_0$ constraints from CF3 -- which combines peculiar velocities from three galaxies -- and from SBF -- which uses only NGC 4993's peculiar velocity -- are tighter than our fiducial result despite using fewer galaxies than our six-galaxy DESI sample. These comparisons raise an important question: Why do DESI results show less precision than previous studies?

For many studies, however, direct comparison is complicated by methodological differences that likely account for much of the apparent difference in constraining power. For example, \citet{gw170817_nature}, \citet{nicolaou}, and \citet{palmese_ag} model peculiar motion by combining catalog galaxy peculiar velocities weighted by their distance from NGC 4993. This approach can assign non-trivial weights to galaxies outside the host group, meaning that some constraining power derives from galaxies that may not be associated with the GW170817 host. In contrast, our analysis restricts the group peculiar velocity measurement to galaxies we confidently identify as group members. We further test (in Section \ref{subsection:posterior}) whether our results remain robust to the inclusion or exclusion of several questionable group members. Without such precautions, previous measurements may underestimate the uncertainty on the standard siren $H_0$ posterior. We further note that the uniform $H_0$ prior we use in this analysis may lead to broader constraints and larger values than the aforementioned analyses as those typically use a flat-in-log prior.

A more straightforward comparison can be made to the results in \citet{standard_siren_speeds} since they carefully select group galaxies for their peculiar velocity sample. The measurement of $H_0 =58.0^{+6.1}_{-5.3}$ km s$^{-1}$ Mpc$^{-1}$ in \citet{standard_siren_speeds} of three group galaxies from \citet{tully_ngc4993_pv} is similar to our measurement since both a) only use galaxies verified to be in the group in the PV sample and b) include the maximal number of group PV measurements available. In fact, we attempt to recover the result using our likelihood formulation (from Section \ref{subsection:posterior}) by using the same three group galaxy log-distance ratios (derived from the distance moduli and fitted cosmology in CF3). We show the resulting posterior in Figure~\ref{fig:{H0_prior_work}}, which is shifted to noticeably higher value of $H_0 = 69.4^{+7.0}_{-6.1}$ km s$^{-1}$ Mpc$^{-1}$. The increase can mostly be attributable to our choice of $z_{\mathrm{tot}}$. We use the brightest group galaxy redshift for $z_{\mathrm{tot}}$ while \citet{standard_siren_speeds} use the redshift of NGC 4993. As shown in Section \ref{subsection:H0}, this choice can have a significant effect on the location of the final $H_0$ posterior. When we use the reported values from \citet{standard_siren_speeds} for $z_{\mathrm{tot}}$ and $\eta_{\mathrm{obs}}$, we confirm that we recover a value of $H_0 = 60.3^{+6.6}_{-6.3}$ km s$^{-1}$ Mpc$^{-1}$ in conformance with that reported in \citet{standard_siren_speeds}.

We make a final comparison to the results of \citet{mukherjee_2021}, which uses the method of velocity reconstruction to determine the peculiar velocity of NGC 4993. Rather than determining the coherent group velocity of the merger's host group, a velocity reconstruction approach extracts the host PV from a broader, three-dimensional velocity map of the universe at the coordinates of the host. This map is constructed by combining the data of galaxy surveys that comprehensively characterize the large scale structure of the universe with an assumed model of gravitational dynamics. Using this technique, \citet{mukherjee_2021} find $v_{\mathrm{pec}}=390 \pm 130$ km s$^{-1}$ and $H_0 = 68.3^{+4.6}_{-4.5}$ km s$^{-1}$ Mpc$^{-1}$. The PV and $H_0$ central values from velocity reconstruction are \textit{both} lower than our fiducial, which is interesting since PVs and $H_0$ are inversely related. In fact, Table \ref{tab:results_samples} shows a few measurements with PVs near that of \citet{mukherjee_2021}, but all of these measurements have higher posterior $H_0$ values. This might point to a difference in modeling the posterior that does not relate to peculiar velocities (such as the question regarding the value of $z_{\mathrm{tot}}$ when comparing to \citealt{standard_siren_speeds}). In any case, this  measurement agrees with our fiducial result at the $1\sigma$ level, demonstrating the consistency of different methodological approaches to determining $H_0$ from standard sirens.

Turning to the comparative $H_0$ precision of the CF3, SBF, and DESI results, the broader DESI constraints simply reflect the larger uncertainties in DESI galaxy peculiar velocities compared to those from CF3 and SBF. As shown in Figure~\ref{fig:{phase-space}}, there is a single galaxy that is in both the CF3 and DESI fiducial FP sample, and the peculiar velocity of this galaxy is measured to similar precision of $\sigma_{\eta_{obs}} \sim 0.1$ in both surveys. This level of precision is generally representative of our uncertainties on FP galaxy log-distance ratios. The other two CF3 galaxies, which we also observed as part of our TF survey, have significantly better precision of $\sigma_{\eta_{obs}} \sim 0.06$. We note that $\sigma_{\eta_{obs}}$ is known to be smaller in CF3 compared to DESI when considering TF galaxies \citep{douglass_tfr_edr}. Given this better precision, it is unsurprising that CF3 provides a better $H_0$ constrain than DESI even though more galaxies are involved in the calculation of the DESI group peculiar velocity. A similar line of reasoning explains why the SBF measurement -- which typically provides better precision than both TF and FP -- provides significantly tighter $H_0$ constraints than that of both CF3 and DESI. Impressively, the SBF precision leads to a 5\% uncertainty on $H_0$.

More than providing tighter $H_0$ constraints, the SBF PV measurement avoids the assumption that NGC 4993 lies in a relaxed group, which underpins the group-averaging method for determining peculiar velocity. In Section \ref{subsection:cluster_mem}, we used optical DESI data to make an initial foray into this topic, but our results provide no conclusive evidence on whether or not the group is relaxed. Although prior studies have noted that X-ray observations suggest a lack of relaxation \citep{hjorth_2017}, we know of no systematic study of this group's virialization status. Such observations might definitively determine the veracity of the relaxation assumption. If this assumption were to fail, it is likely that the dominant component of NGC 4993's peculiar motion would not necessarily be attributable to its broader group motion. Under such a circumstance, our measurement of $\eta_{\mathrm{obs}}$ would inaccurately reflect the log-distance ratio of NGC 4993 even if our individual galaxy log-distance ratio measurements were accurate. One would be required to only use log-distance ratio information from NGC 4993 itself (rather than its group neighbors) to measure the true log-distance ratio of the standard siren event. We in fact utilize this ``host-galaxy-based'' approach when using $\eta_{\mathrm{obs}}$ from a single SBF measurement of NGC 4993. A further benefit of only measuring the log-distance ratio of NGC 4993 relates to the question of whether to use the group or NGC 4993 redshift for $z_{\mathrm{tot}}$. This issue becomes irrelevant when only measuring the log-distance ratio of NGC 4993. Since both the luminosity distance and peculiar velocity are directly measured from the merger galaxy host, it is unambiguous that the value of $z_{\mathrm{tot}}$ should the galaxy host redshift. A final benefit to only using one peculiar velocity measurement relates to  covariance between log-distance ratios calculated using the same FP calibration. In our DESI sample, for example, there is a covariance between all FP log-distance ratios since they all derive from the same FP calibration. Given the magnitude of our uncertainties on log-distance ratios, we have been justified in neglecting this covariance. If, however, these covariances were non-negligible, accounting for them would make the computation of Equation \ref{eq:H0_post} more complex. When measuring only a single log-distance ratio, the issue of covariance between measurements disappears, adding another great benefit to the ``host-galaxy-based'' approach. 

As a last note, we turn to a systematic that lurks in all standard siren measurements that account for peculiar velocities via direct measurement. In Section \ref{subsection:zero_point}, we fixed the PV zero-point using a sample of Type Ia supernovae distances. This set of distances conforms to a value of $H_0$ derived via the distance ladder used in SH0ES analysis. Thus, in a sense, our standard siren measurement conforms to a distance ladder $H_0$ measurement. Since the primary benefit of standard siren analyses is that they should be unaffected by the same systematics that might plague distance ladder measurements, the fact that PV measurements utilize distance ladders tarnishes the claim that standard sirens $H_0$ measurements serve as an independent check on distance ladder $H_0$ measurements. For direct measurements of PVs, this reality is unavoidable so long as the peculiar velocity of the galaxy host is non-negligible. If new nearby standard siren events are detected and the standard siren precision improves, this systematic may make standard sirens less of an ``independent check'' on the measurement of $H_0$. It is worth noting that future DESI standard siren analyses will be calibrated following other calibrators beyond Supernovae, as the calibration of the DESI Data Release 2 (DR2) PV analyses will allow for larger statistics for the calibration sample and hence a larger overlap with other distance indicators, which can be used to produce PV measurements independent of Supernovae. Once those calibrations are available, the results of this work can also be updated. For what concerns the currently available calibration from \citet{DESI_zero_point}, we note that the SBF-based calibration yields results that are consistent with the fiducial result, which is expected since in both cases the distance ladder is Cepheid-calibrated. In other words, it is reasonable to assume that our result is robust to different choices of the third distance ladder rung, whether that is Supernovae or SBF, but relies on Cepheid distances.  

It is important to note that this distance ladder dependence is not expected to affect velocity reconstruction approaches to standard siren peculiar velocities. In fact, \citet{mukherjee_2021} note that their velocity reconstruction method -- which assumes an $H_0$ value -- is mostly insensitive to the choice of this value.

\section{Conclusion}\label{section:Conclusion}
Many bright standard siren measurements have been made using GW event GW170817 and its host galaxy, NGC 4993, to derive a value for $H_0$. When EM observations of the jet and its afterglow enable stronger constraints on the luminosity distance, the peculiar velocity can become the dominant uncertainty on the measurement of $H_0$ \citep{he_2019, nimonkar_2024}. We have performed a dedicated set of DESI observations to explore the peculiar velocities of galaxies in NGC 4993's group to better constrain the group PV and $H_0$. Our fiducial sample includes peculiar velocities measured from six Fundamental Plane galaxies using the DESI DR1 peculiar velocity calibration. We also measure peculiar velocities of eight Tully-Fisher galaxies, which are excluded from our sample due to zero-pointing complications. Our galaxy samples were selected using the same photometric selection criteria used in the DESI Peculiar Velocity Survey \citep{target_selection}. Our fiducial sample has double the number of group members contributing to the PV sample compared to previous studies, yet our $H_0$ constraint shows little improvement due to large peculiar velocity uncertainties on individual galaxies. All of our DESI results demonstrate good agreement with both \emph{Planck} and SH0ES $H_0$ measurements at the $2\sigma$ level. Using a single (non-DESI) SBF measurement of the peculiar velocity of NGC 4993 alone, we are able to place the tightest standard siren constraint on $H_0$ to date at 5\%. Since our results demonstrate the galaxy group may not be virialized, this approach is preferable as it avoids the problematic assumption of virialization.

In the near future, we expect tens of GW events to be accompanied by a detectable kilonova \citep[e.g.][]{Kunnumkai_GW230529, kunnumkai2024_O4} or even a jet afterglow \citep[e.g.][]{Kaur_2024}, leading to 2\% level precision on the Hubble constant \citep{2023ApJ...958..158K}. Careful estimates of the host peculiar velocity may be even more crucial when combining a number of nearby multimessenger events to reach that level of precision with accuracy. Overall, this work shows the promise of multi-object spectroscopic observations by DESI to provide insights on standard siren $H_0$ constraints, while also probing the environments around compact object mergers \citep{2019BAAS...51c.310P,bin_host_con}.

\section*{Data Availability}
All data shown in the above figures will be available on Zenodo upon acceptance.

\section*{Acknowledgements}
\noindent AA and AP acknowledge support for this work by NSF Grant No. 2308193. This material is based upon work supported by the U.S. Department of Energy (DOE), Office of Science, Office of High-Energy Physics, under Contract No. DE–AC02–05CH11231, and by the National Energy Research Scientific Computing Center, a DOE Office of Science User Facility under the same contract. Additional support for DESI was provided by the U.S. National Science Foundation (NSF), Division of Astronomical Sciences under Contract No. AST-0950945 to the NSF’s National Optical-Infrared Astronomy Research Laboratory; the Science and Technology Facilities Council of the United Kingdom; the Gordon and Betty Moore Foundation; the Heising-Simons Foundation; the French Alternative Energies and Atomic Energy Commission (CEA); the National Council of Humanities, Science and Technology of Mexico (CONAHCYT); the Ministry of Science, Innovation and Universities of Spain (MICIU/AEI/10.13039/501100011033), and by the DESI Member Institutions: \url{https://www.desi.lbl.gov/collaborating-institutions}.

The DESI Legacy Imaging Surveys consist of three individual and complementary projects: the Dark Energy Camera Legacy Survey (DECaLS), the Beijing-Arizona Sky Survey (BASS), and the Mayall z-band Legacy Survey (MzLS). DECaLS, BASS and MzLS together include data obtained, respectively, at the Blanco telescope, Cerro Tololo Inter-American Observatory, NSF’s NOIRLab; the Bok telescope, Steward Observatory, University of Arizona; and the Mayall telescope, Kitt Peak National Observatory, NOIRLab. NOIRLab is operated by the Association of Universities for Research in Astronomy (AURA) under a cooperative agreement with the National Science Foundation. Pipeline processing and analyses of the data were supported by NOIRLab and the Lawrence Berkeley National Laboratory. Legacy Surveys also uses data products from the Near-Earth Object Wide-field Infrared Survey Explorer (NEOWISE), a project of the Jet Propulsion Laboratory/California Institute of Technology, funded by the National Aeronautics and Space Administration. Legacy Surveys was supported by: the Director, Office of Science, Office of High Energy Physics of the U.S. Department of Energy; the National Energy Research Scientific Computing Center, a DOE Office of Science User Facility; the U.S. National Science Foundation, Division of Astronomical Sciences; the National Astronomical Observatories of China, the Chinese Academy of Sciences and the Chinese National Natural Science Foundation. LBNL is managed by the Regents of the University of California under contract to the U.S. Department of Energy. The complete acknowledgments can be found at \url{https://www.legacysurvey.org/}.

Any opinions, findings, and conclusions or recommendations expressed in this material are those of the author(s) and do not necessarily reflect the views of the U. S. National Science Foundation, the U. S. Department of Energy, or any of the listed funding agencies.

The authors are honored to be permitted to conduct scientific research on Iolkam Du’ag (Kitt Peak), a mountain with particular significance to the Tohono O’odham Nation.

This research has made use of data or software obtained from the Gravitational Wave Open Science Center (gwosc.org), a service of the LIGO Scientific Collaboration, the Virgo Collaboration, and KAGRA. This material is based upon work supported by NSF's LIGO Laboratory which is a major facility fully funded by the National Science Foundation, as well as the Science and Technology Facilities Council (STFC) of the United Kingdom, the Max-Planck-Society (MPS), and the State of Niedersachsen/Germany for support of the construction of Advanced LIGO and construction and operation of the GEO600 detector. Additional support for Advanced LIGO was provided by the Australian Research Council. Virgo is funded, through the European Gravitational Observatory (EGO), by the French Centre National de Recherche Scientifique (CNRS), the Italian Istituto Nazionale di Fisica Nucleare (INFN) and the Dutch Nikhef, with contributions by institutions from Belgium, Germany, Greece, Hungary, Ireland, Japan, Monaco, Poland, Portugal, Spain. KAGRA is supported by Ministry of Education, Culture, Sports, Science and Technology (MEXT), Japan Society for the Promotion of Science (JSPS) in Japan; National Research Foundation (NRF) and Ministry of Science and ICT (MSIT) in Korea; Academia Sinica (AS) and National Science and Technology Council (NSTC) in Taiwan.
The authors are grateful for computational resources provided by the LIGO Laboratory and supported by National Science Foundation Grants PHY-0757058 and PHY-0823459.
\bibliographystyle{yahapj}
\bibliography{references1}

\end{document}